\documentclass[sigconf]{acmart}

\AtBeginDocument{%
  }

\acmPrice{15.00}
\acmISBN{978-1-4503-XXXX-X/18/06}
\copyrightyear{2024}
\acmYear{2024}
\setcopyright{rightsretained}
\acmConference[CHI '24]{Proceedings of the CHI Conference on Human Factors in Computing Systems}{May 11--16, 2024}{Honolulu, HI, USA}
\acmBooktitle{Proceedings of the CHI Conference on Human Factors in Computing Systems (CHI '24), May 11--16, 2024, Honolulu, HI, USA}
\acmDOI{10.1145/3613904.3642867}
\acmISBN{979-8-4007-0330-0/24/05}

\usepackage{acmart-taps}
\usepackage{graphicx}
\usepackage{multirow}
\usepackage{url}

\newcommand{\sysname}[0]{VIVID}

\begin{document}

\title{\sysname{}: Human-AI Collaborative Authoring of Vicarious Dialogues from Lecture Videos}

\author{Seulgi Choi}
\affiliation{%
  \institution{School of Computing, KAIST}
  \city{Daejeon}
  \country{Republic of Korea}}
  \email{igules8925@kaist.ac.kr}

\author{Hyewon Lee}
\affiliation{%
  \institution{School of Computing, KAIST}
  \city{Daejeon}
  \country{Republic of Korea}}
\email{hyewon0809@kaist.ac.kr}

\author{Yoonjoo Lee}
\affiliation{%
  \institution{School of Computing, KAIST}
  \city{Daejeon}
  \country{Republic of Korea}}
\email{yoonjoo.lee@kaist.ac.kr}

\author{Juho Kim}
\affiliation{%
  \institution{School of Computing, KAIST}
  \city{Daejeon}
  \country{Republic of Korea}}
\email{juhokim@kaist.ac.kr}

\renewcommand{\shortauthors}{Seulgi Choi, Hyewon Lee, Yoonjoo Lee, Juho Kim}

\begin{abstract}
The lengthy monologue-style online lectures cause learners to lose engagement easily. Designing lectures in a “vicarious dialogue” format can foster learners’ cognitive activities more than monologue-style. However, designing online lectures in a dialogue style catered to the diverse needs of learners is laborious for instructors. We conducted a design workshop with eight educational experts and seven instructors to present key guidelines and the potential use of large language models (LLM) to transform a monologue lecture script into pedagogically meaningful dialogue. Applying these design guidelines, we created \sysname{} which allows instructors to collaborate with LLMs to design, evaluate, and modify pedagogical dialogues. In a within-subjects study with instructors (N=12), we show that \sysname{} helped instructors select and revise dialogues efficiently, thereby supporting the authoring of quality dialogues. Our findings demonstrate the potential of LLMs to assist instructors with creating high-quality educational dialogues across various learning stages.
\end{abstract}



\maketitle

\section{Introduction}
Online lectures are widely used for conveying knowledge in various learning contexts. Notably, instructors often use them as educational resources in various teaching contexts like flipped learning ~\cite{lee2017development} or supplementary materials~\cite{brecht2012learning,ellman2016article}, which usually take the format of knowledge-transfer-oriented online lectures. However, they usually take the form of a lengthy monologue. This format can cause learners to feel disengaged or quickly lose interest~\cite{beheshti2018characteristics}, potentially resulting in persistent negative emotions that detrimentally affect learning outcomes~\cite{graesser2012emotions}. To address the limitations of this lecture format, studies have explored the application of conversational agents (CA) to video-based learning ~\cite{nugraha2020tool,tanprasert2022authoring,tanprasert2023scripted,winkler2020sara}. Many studies used CAs to mimic human tutoring behaviors such as scaffolding ~\cite{graesser1999autotutor}, and these direct interactions with CA have improved the learning experience of online learners. 

Although these studies imply the importance of CAs' scaffolding mechanisms in online video lecture settings, they have supported mostly the learners who prefer to interact directly with an instructor and peer learners~\cite{Sutton2001ThePO,swan2003learning}. Yet, for \textit{vicarious learners}, who prefer to learn from others and actively process the interactions of others, interactions that can be vicariously processed are more beneficial to their learning~\cite{Sutton2001ThePO,swan2003learning}. To enhance \textit{vicarious learners}' experience, systems with multiple CAs that simulate interactions between an instructor and a \textit{direct learner}~\cite{tanprasert2023scripted} have been introduced based on \textit{vicarious learning theory}~\cite{Mayes2015StillTL}. \textit{Vicarious learning theory} explains the benefits of learning when \textit{vicarious learners} observe tutoring between an instructor and \textit{a direct learner} who interact directly with the instructor in a video lecture~\cite{chi2016why,chi2008observing,Mayes2015StillTL,Bandura1977SocialLT}. The studies found that \textit{vicarious learners} preferred dialogic lecture videos that incorporate CAs over monologue-style lecture videos, and it resulted in a positive effect on students' engagement~\cite{tanprasert2023scripted}. Therefore, introducing vicarious dialogue into monologue-style lectures can serve as a promising solution to address the limitations of conventional online lectures and satisfy \textit{vicarious learners}. 

However, the current approach has not yet addressed how to create high-quality dialogues that cater to vicarious learners through adaptation or expansion of original lecture contents. It is important to consider the quality of the learning content, such as the level of detail provided by the lecturer, because it can significantly affect a vicarious learner's cognitive load and engagement. Thus, rather than simply enhancing lectures, we converted the original lecture script into a format that can reduce the cognitive load for vicarious learners and found a pedagogically meaningful format for high-quality dialogue. To do this, we integrated LLM in this conversion process since LLM has been discussed as a feasible way to design vicarious dialogues while reducing the extra effort for instructors~\cite{tanprasert2023scripted}.
In short, this work aims to alleviate the manual effort of instructors authoring vicarious dialogue and establish a scalable pipeline for designing educationally high-quality dialogues from lectures.

To achieve this goal, we have developed five guidelines for transforming monologic lectures into a vicarious dialogue that can benefit online learners: \textit{Dynamic, Academically Productive, Cognitive Adaptable, Purposeful}, and \textit{Immersive}. As an initial step in crafting these guidelines, we conducted an iterative inductive literature analysis to define what constitutes a pedagogically meaningful dialogue. However, most existing literature focused on insights derived from classrooms or intelligent tutoring systems, not video lectures. Furthermore, there is limited research on transforming the content in video lectures into high-quality educational dialogues. Therefore, we conducted a design workshop with eight educational experts and seven secondary school teachers to develop the guidelines to be tailored for a STEM video learning setting. 

To facilitate the efficient authoring of video-based vicarious dialogues based on our guidelines, we propose a system, \sysname{} (\textbf{VI}deo to \textbf{VI}carious \textbf{D}ialogue), which allows instructors to design, evaluate, and modify vicarious interactions with video lectures. To empower this system, we propose a collaborative design process between LLM and instructors to generate high-quality vicarious dialogues efficiently. This process consists of three stages, guided by the developed guidelines in the workshop: (1) \textbf{Initial Generation}: After an instructor chooses where to convert in a lecture, LLM configures a direct learner's understanding level for each concept in the selected section of the lecture and generates initial dialogues.
(2) \textbf{Compare and Selection}: Instructors compare and select from multiple generated dialogues, and (3) \textbf{Refinement}: Instructors collaborate with LLM to refine the final dialogue, which will replace a section of the video lecture. 

To determine whether \sysname{} is helpful for instructors to transform monologue lectures into high-quality dialogue lectures, we conducted a within-subjects study with 12 instructors. \sysname{} helped instructors simulate a direct learner effectively through co-designing with \sysname{}. Furthermore, instructors found that VIVID is significantly better in monitoring essential considerations ($p=0.04$)  with an effect size (Cohen’s d) of 0.8 than the Baseline when designing dialogue.
To evaluate the pedagogical quality of the authored dialogues designed through \sysname{}, we also conducted a human evaluation with six secondary instructors in four criteria which is if the dialogue is \textit{Dynamic}, \textit{Academically productive,} \textit{Immersive}, and \textit{Correct}. We found that the dialogues made by \sysname{} were significantly better quality in most criteria than the dialogues generated by Baseline.

The contributions of this work are as follows:
\begin{itemize}
    \item Design guidelines through design workshop for making vicarious educational dialogues from lecture videos. 
    \item \sysname{}, a system that collaborates with LLM to assist instructors in authoring vicarious dialogues from the monologue-styled lecture videos.
    \item Findings from a user study with 12 instructors showing how \sysname{} can assist instructors in dialogue authoring (Section 6.2), and a technical evaluation with six instructors that demonstrates the higher quality of dialogues created by instructors using \sysname{} compared to the Baseline (Section 6.4).
\end{itemize}

\section{Related Work}
We reviewed previous research on simulating vicarious learning in online learning contexts and approaches for generating diverse educational dialogues at scale.
\subsection{Simulating Vicarious Learning in an Online Learning Environment.}
Vicarious learning~\cite{chi2016why,ding2021learning} in an online environment typically occurs when observing the interaction between other learners and an instructor on platforms like Zoom or when witnessing peer discussions on QA platforms. Such situations of vicarious learning can stimulate learners’ cognitive activity and enhance their level of engagement. 

Thus, research has employed a Conversational Agent (CA) ~\cite{ruan2019quizbot,fast2018iris,grossman2019mathbot,wambsganss2021arguetutor,lee2021curiosity} to simulate interactions between a virtual tutor and tutee for supporting vicarious learners in video-based learning. For instance, Nugraha et al. ~\cite{nugraha2020tool} explored how a CA in the role of a tutee to Massive Open Online Course (MOOC) videos could enhance the vicarious learners' learning experiences. Similarly, Tanprasert et al. ~\cite{tanprasert2022authoring} implemented vicarious interaction in MOOCs as if participating in a Zoom class. To do this, they added scripted vicarious dialogues between virtual learners and an instructor to a lecture video in a chat format. These studies showed learners preferred dialogue-like lecture videos with CAs that mimic vicarious interactions over monologue lecture videos. This type of interpersonal interaction positively impacted vicarious learners' engagement.

However, it's important to note that these studies employed manually crafted dialogues of assumed equal quality even though the quality of dialogues can significantly influence learner engagement and outcomes~\cite{paladines2020systematic}. Furthermore, there is limited research on designing high-quality educational dialogues to facilitate vicarious learning in video-based learning contexts. Consequently, we aim to fill this research gap by developing guidelines for creating high-quality educational dialogues that can promote effective vicarious learning experiences~\cite{tanprasert2023scripted}.

\subsection{Generating Diverse Educational Dialogues for Vicarious Learners at Scale} 

Large Language Models (LLMs) are becoming increasingly useful for educators~\cite{martinez2014mtfeedback,wang2021seeing}. One promising area of research involves utilizing them to create a wide range of educational materials~\cite{denny2023human,sarsa2022automatic}. For example, Wang et al. ~\cite{wang2018qg} found that large pretrained language models (PLMs) can automate the generation of educational assessment questions.
Other approaches introduce question generation models that automatically produce questions from educational content such as textbooks~\cite{wang2018qg, willis2019key}. 

However, they primarily focus on addressing the challenge of scaling the generation of specific question types and provide solutions primarily at the model level without considering the needs of instructors and learners. In contrast, Promptiverse~\cite{lee2022promptiverse} proposes a novel approach aimed at reducing the workload for instructors while delivering useful and diverse prompts to learners. Furthermore, ReadingQuizMaker~\cite{lu2023readingquizmaker} introduces a system to enable instructors to conveniently generate high-quality questions. Both systems allow instructors to create prompts or questions at scale, but neither considers the learners' level when generating them. Furthermore, they mainly focus on enhancing the diversity of single prompts or quizzes. Consequently, applying these approaches to generating diverse educational dialogues, which involve dynamic interactions between tutees and a tutor, may present challenges.  

To evaluate various uses of LLM in generating learning materials, such as code explanations~\cite{Leinonen2023ComparingCE}, learning objectives~\cite{sridhar2023harnessing}, they have been evaluated based on general criteria, such as ``easy to understand'' or ``accuracy'' without thoroughly considering the quality for specific tasks. 
However, to ensure quality, it is essential to establish specific and measurable criteria tailored to each task.
Moreover, integrating LLM into education practice requires balancing the use of LLM with the role of instructors since relying solely on an automatic pipeline with LLM may result in low quality. Thus, we aim to establish criteria for assessing education dialogues and propose an LLM-based pipeline that can generate high-quality dialogues scalable while considering vicarious learners. Further, based on this pipeline, we aim to design an interactive system that allows collaboration between LLM and instructors in authoring dialogues.

\section{Design Workshop}
To develop a guideline for designing high-quality vicarious dialogues, we employed the two-step approach. In the first step, we conducted an iterative inductive literature analysis to define what a pedagogically meaningful dialogue should look like. 
Despite the increasing amount of research on video learning, there has been little research on how to design beneficial vicarious dialogues based on lecture videos and how to support instructors in doing this.
To address these issues, we conducted a design workshop to develop new guidelines for designing vicarious dialogues in the context of video-based learning.

\subsection{Utterance Patterns and Teaching Strategies}
Two of the authors conducted an iterative inductive analysis of literature to define what constitutes a pedagogically meaningful dialogue in literature. To identify relevant literature, we conducted a query-based search with the PRISMA process~\cite{moher2009} on Google Scholar and the ACM Digital Library, and the 50 final papers were selected for meta-analysis.
Our review was based on three search queries related to main keywords (Detailed analysis method is in the Supplemental Material):
\begin{itemize}
    \item \textbf{Vicarious Learning:} "vicarious learning" + ("learning gain" OR "tutorial dialogue" OR "monologue")
    \item \textbf{Classroom Interaction:} "classroom interaction"  + "science" + "dialogic" + "teacher questioning" + ("secondary school" OR "undergraduate")
    \item \textbf{Human Tutoring:} "human tutoring" + "tutorial dialogue" + ("strategy" OR "move") +("scaffolding" OR "feedback")
\end{itemize}

Based on our literature analysis, we created initial guidelines for designing vicarious dialogue in video lectures. The vicarious dialogue should be perceived as a natural conversation occurring during a lecture and should be effective for vicarious learners. Thus, we derived two main factors for designing vicarious dialogues: (1) the most commonly observed utterance categories in real tutoring (Table~\ref{tab:tutor_category}, Table~\ref{tab:tutee_category}) and (2) effective teaching strategies for vicarious learners.

\subsubsection{Key utterance categories that are commonly observed in 1-to-1 tutoring and classroom.}

Several studies have collected data from actual one-on-one tutoring or classroom session recordings and performed qualitative coding at the statement level to classify representative types of utterances made by tutors and tutees. We categorized the tutor's utterances into nine types (Table~\ref{tab:tutor_category}) and the learner's utterances into five types (Table~\ref{tab:tutee_category}) to utilize for designing vicarious dialogues that simulate a natural tutoring scenario.

\subsubsection{Three teaching strategies that can positively affect vicarious learners.}
Research indicates that vicarious learners are notably affected by the direct learner's discourse following the instructor's statements as vicarious learners tend to mimic direct learner's actions~\cite{chi2016why}.
The three most influential dialogue patterns in vicarious learning include:

\textbf{\textit{Integrate a direct learner's cognitive conflict.}}
A tutoring video that contains a \textit{cognitive conflict} situation, where the instructor corrects errors made by the direct learner, can improve the attention and interest of vicarious learners~\cite{ding2021learning, muldner2011learning,chi2016why}. Thus, we propose designing dialogues as if the instructor encourages the direct learner to reach confusion and addresses misconceptions productively~\cite{Lehman2012ConfusionAC,VanLehn2003WhyDO,Graesser2004AutoTutorAT}.

\textbf{\textit{Integrate a direct learner's deep-level reasoning questions.}} We suggest incorporating deep-level reasoning questions ~\cite{driscoll2003vicarious,craig2006deep,gholson2009exploring,craig2012promoting,Craig2000OverhearingDA} that address comparisons, inferences, and causal relationships among concepts into the direct learner's utterances. According to previous research in Intelligent Tutoring System (ITS)~\cite{driscoll2003vicarious, craig2006deep, gholson2009exploring, craig2012promoting,Craig2000OverhearingDA}, the vicarious learners' learning was significantly improved when the direct learner posed deep questions. Therefore, if a direct learner asks a deep-level reasoning question during a lecture, it can encourage vicarious learners to engage in critical thinking.

\textbf{\textit{Integrate a direct learner's substantial and relevant follow-up responses.}} A direct learner should provide answers or self-explanations based on the learning contents followed by an instructor’s scaffolding or lecturing~\cite{craig2012promoting, chi2008observing, chi2016why,Dubovi2019InstructionalSF}.

\aptLtoX{\begin{table*}[!h]
\begin{tabular}{l|l|l|l}
\hline
\textbf{Tutor's utterance} & \textbf{Definition}                                                                                                                                                                                          & \textbf{Tutor's utterance}       & \textbf{Definition}                                                                                                                                  \\ \hline
Self-monitoring            & \begin{tabular}[c]{@{}l@{}}Utterance related to self-monitoring of one's teaching style ~\cite{Chi2001LearningFH}.\end{tabular}                                                                                                      & Summarizing                      & \begin{tabular}[c]{@{}l@{}}Utterance that summarizes what has bee done so far or restates student's questions/statements/ comments~\cite{Cade2008DialogueMI,Lu2009ExpertVN,Bansal2018TeacherDM,Lyle2008DialogicTD,Ong2016PromotingHT,Kranzfelder2019TheCD,Litman2006CorrelationsBD}\end{tabular} \\ \hline
Lecturing                  & \begin{tabular}[c]{@{}l@{}}Utterance explaining declarative knowledge, which includes facts and conceptual principles.  ~\cite{Chi2001LearningFH,Cade2008DialogueMI,Lu2009ExpertVN,Boyer2009DiscoveringTD,Kranzfelder2019TheCD,Brodie2011WorkingWL}\end{tabular}                                                                        & {Answering} & Utterance in response to student questions~\cite{Chi2001LearningFH,Lu2009ExpertVN,Bansal2018TeacherDM,Teo2016ExploringTD,Litman2006CorrelationsBD}.                                                                                                         \\ \hline
Demonstrating              & \begin{tabular}[c]{@{}l@{}}Utterance related to solving specific problems in a way that allows student to model the instructor's problem-solving approach ~\cite{Cade2008DialogueMI,Lu2009ExpertVN}.\end{tabular}                                  & Scaffolding                      & \begin{tabular}[c]{@{}l@{}}Utterance involving assistance or hints to help students reach answers on their own. ~\cite{Chi2001LearningFH,Cade2008DialogueMI,Lu2009ExpertVN,Boyer2009DiscoveringTD,Muhonen2016ScaffoldingTD,Chin2006ClassroomII,Bansal2018TeacherDM,Lyle2008DialogicTD,Ong2016PromotingHT,Muhonen2017QualityOE,Kranzfelder2019TheCD,Brodie2011WorkingWL,Hume1996HintingAA,Litman2006CorrelationsBD,Mitrovic2013TheEO,DMello2010ExpertTF,Boyer2008BalancingCA,boyer2008learner}\end{tabular}                       \\ \hline
Questioning                & \begin{tabular}[c]{@{}l@{}}Utterance containing questions to encourage students to recall knowledge or think productively  (e.g., deep-level reasoning/short-answer questions ~\cite{Chi2001LearningFH,Vrikki2019DialogicPI,Chin2006ClassroomII,Bansal2018TeacherDM,Lyle2008DialogicTD,Ong2016PromotingHT,Muhonen2017QualityOE,Teo2016ExploringTD,Kranzfelder2019TheCD,LevyS2013ExaminingSO,Brodie2011WorkingWL,Hume1996HintingAA,boyer2008learner,Litman2006CorrelationsBD,Boyer2009DiscoveringTD,Ros2003TheRO}.\end{tabular} & Diagnosing                       & \begin{tabular}[c]{@{}l@{}}Utterance used to diagnose student understanding  or progress~\cite{Chi2001LearningFH,Lu2009ExpertVN,Boyer2009DiscoveringTD}.\end{tabular}                                             \\ \hline
Off-topic                  & \begin{tabular}[c]{@{}l@{}}Introduction or unrelated utterances, such as  small talk not related to learning ~\cite{Cade2008DialogueMI,Lu2009ExpertVN,Teo2016ExploringTD,Litman2006CorrelationsBD}.\end{tabular}                                                                                 &                                  &        \\ \hline
\end{tabular}%
\caption{This table displays nine categories of tutor utterances and their corresponding definitions. The table consists of two columns, the first containing the tutor's utterance categories, and the second containing their respective definitions. Categories include Self-monitoring, Lecturing, Demonstrating, Questioning, Off-topic, Summarizing, Answering, Scaffolding, and Diagnosing.}
\label{tab:tutor_category}
\end{table*}}{\begin{table*}[!h]
\resizebox{\textwidth}{!}{%
\begin{tabular}{l|l|l|l}
\hline
\textbf{Tutor's utterance} & \textbf{Definition}                                                                                                                                                                                          & \textbf{Tutor's utterance}       & \textbf{Definition}                                                                                                                                  \\ \hline
Self-monitoring            & \begin{tabular}[c]{@{}l@{}}Utterance related to self-monitoring of one's\\ teaching style ~\cite{Chi2001LearningFH}.\end{tabular}                                                                                                      & Summarizing                      & \begin{tabular}[c]{@{}l@{}}Utterance that summarizes what has bee done\\ so far or restates student's questions/statements/\\ comments~\cite{Cade2008DialogueMI,Lu2009ExpertVN,Bansal2018TeacherDM,Lyle2008DialogicTD,Ong2016PromotingHT,Kranzfelder2019TheCD,Litman2006CorrelationsBD}\end{tabular} \\ \hline
Lecturing                  & \begin{tabular}[c]{@{}l@{}}Utterance explaining declarative knowledge,\\ which includes facts and conceptual principles. \\ ~\cite{Chi2001LearningFH,Cade2008DialogueMI,Lu2009ExpertVN,Boyer2009DiscoveringTD,Kranzfelder2019TheCD,Brodie2011WorkingWL}\end{tabular}                                                                        & {Answering} & Utterance in response to student questions~\cite{Chi2001LearningFH,Lu2009ExpertVN,Bansal2018TeacherDM,Teo2016ExploringTD,Litman2006CorrelationsBD}.                                                                                                         \\ \hline
Demonstrating              & \begin{tabular}[c]{@{}l@{}}Utterance related to solving specific problems\\ in a way that allows student to model the\\ instructor's problem-solving approach ~\cite{Cade2008DialogueMI,Lu2009ExpertVN}.\end{tabular}                                  & Scaffolding                      & \begin{tabular}[c]{@{}l@{}}Utterance involving assistance or hints to\\ help students reach answers on their own. \\ ~\cite{Chi2001LearningFH,Cade2008DialogueMI,Lu2009ExpertVN,Boyer2009DiscoveringTD,Muhonen2016ScaffoldingTD,Chin2006ClassroomII,Bansal2018TeacherDM,Lyle2008DialogicTD,Ong2016PromotingHT,Muhonen2017QualityOE,Kranzfelder2019TheCD,Brodie2011WorkingWL,Hume1996HintingAA,Litman2006CorrelationsBD,Mitrovic2013TheEO,DMello2010ExpertTF,Boyer2008BalancingCA,boyer2008learner}\end{tabular}                       \\ \hline
Questioning                & \begin{tabular}[c]{@{}l@{}}Utterance containing questions to encourage\\ students to recall knowledge or think productively\\ (e.g., deep-level reasoning/short-answer questions\\ ~\cite{Chi2001LearningFH,Vrikki2019DialogicPI,Chin2006ClassroomII,Bansal2018TeacherDM,Lyle2008DialogicTD,Ong2016PromotingHT,Muhonen2017QualityOE,Teo2016ExploringTD,Kranzfelder2019TheCD,LevyS2013ExaminingSO,Brodie2011WorkingWL,Hume1996HintingAA,boyer2008learner,Litman2006CorrelationsBD,Boyer2009DiscoveringTD,Ros2003TheRO}.\end{tabular} & Diagnosing                       & \begin{tabular}[c]{@{}l@{}}Utterance used to diagnose student understanding \\ or progress~\cite{Chi2001LearningFH,Lu2009ExpertVN,Boyer2009DiscoveringTD}.\end{tabular}                                             \\ \hline
Off-topic                  & \begin{tabular}[c]{@{}l@{}}Introduction or unrelated utterances, such as \\ small talk not related to learning ~\cite{Cade2008DialogueMI,Lu2009ExpertVN,Teo2016ExploringTD,Litman2006CorrelationsBD}.\end{tabular}                                                                                 &                                  &                                                                                                                                                      \\ \hline
\end{tabular}%
}
\caption{Nine categories of tutor utterances and their corresponding definitions.}
\label{tab:tutor_category}
\end{table*}}

\aptLtoX{\begin{table*}[!h]
\begin{tabular}{l|l}
\hline
\textbf{Tutee's utterance} & \textbf{Definition}                                                                                                                                                                                           \\ \hline
Questioning                & Utterance related to posing cognitive deep questions or simple questions to the instructor~\cite{Lu2009ExpertVN,Chi2001LearningFH}.                                                                                                                   \\
Answering                  & Utterance related to providing responses or completing scaffolding in response to a instructor's question~\cite{Lu2009ExpertVN,Chi2001LearningFH,Litman2006CorrelationsBD}.                                                                                                    \\
Reflecting                 & Utterance related to assessing one's understanding level in response to a instructor's question or voluntarily~\cite{Lu2009ExpertVN,Chi2001LearningFH}.                                                                                               \\
Explanation                & \begin{tabular}[c]{@{}l@{}}Utterance related to speaking spontaneously, as if articulating one's thoughts simultaneously, without  necessarily being prompted by the instructor's scaffolding~\cite{Lu2009ExpertVN,Chi2001LearningFH}.\end{tabular} \\
Off-topic                  & Introduction or unrelated utterances, such as small talk not related to learning~\cite{Lu2009ExpertVN,Litman2006CorrelationsBD}.                                                                                                                             \\ \hline
\end{tabular}%
\Description{This table presents five categories of tutee utterances and their respective definitions. The table includes two columns: the first column lists the tutee's utterance categories, which consist of Questioning, Answering, Reflecting, Explanation, and Off-topic; the second column provides the definitions of each category. The table is titled 'Table 2: Five categories of tutee utterances and their definitions'.}
\caption{Five categories of tutor utterances and their definitions.}
\label{tab:tutee_category}
\end{table*}}{\begin{table*}[!h]
\resizebox{\textwidth}{!}{%
\begin{tabular}{l|l}
\hline
\textbf{Tutee's utterance} & \textbf{Definition}                                                                                                                                                                                           \\ \hline
Questioning                & Utterance related to posing cognitive deep questions or simple questions to the instructor~\cite{Lu2009ExpertVN,Chi2001LearningFH}.                                                                                                                   \\
Answering                  & Utterance related to providing responses or completing scaffolding in response to a instructor's question~\cite{Lu2009ExpertVN,Chi2001LearningFH,Litman2006CorrelationsBD}.                                                                                                    \\
Reflecting                 & Utterance related to assessing one's understanding level in response to a instructor's question or voluntarily~\cite{Lu2009ExpertVN,Chi2001LearningFH}.                                                                                               \\
Explanation                & \begin{tabular}[c]{@{}l@{}}Utterance related to speaking spontaneously, as if articulating one's thoughts simultaneously, without \\ necessarily being prompted by the instructor's scaffolding~\cite{Lu2009ExpertVN,Chi2001LearningFH}.\end{tabular} \\
Off-topic                  & Introduction or unrelated utterances, such as small talk not related to learning~\cite{Lu2009ExpertVN,Litman2006CorrelationsBD}.                                                                                                                             \\ \hline
\end{tabular}%
}
\Description{This table presents five categories of tutee utterances and their respective definitions. The table includes two columns: the first column lists the tutee's utterance categories, which consist of Questioning, Answering, Reflecting, Explanation, and Off-topic; the second column provides the definitions of each category. The table is titled 'Table 2: Five categories of tutee utterances and their definitions'.}
\caption{Five categories of tutee utterances and their corresponding definitions.}
\label{tab:tutee_category}
\end{table*}}

\subsection{Workshop Overview}
To verify and improve the literature-based guidelines for vicarious interactions in video learning environments, we conducted design workshops with eight educational experts (7 female, 1 male) and seven secondary school teachers (5 female, 2 male). We aimed to (1) derive design guidelines for effective conversion of monologue-style lecture videos into dialogue-style videos and (2) discover design opportunities for a system that can facilitate easy authoring of dialogue-style lectures with LLM. 
We mainly target STEM lectures in our workshop. STEM lectures can cause more intrinsic cognitive load, require more critical thinking than other subjects, and be prone to disengagement while watching lectures because they mostly consist of abstract concepts and complex formulas~\cite{9678899,suhirman2023overcoming}. Thus, we decided to present STEM lectures in a dialogue format as it might help with processing the dense knowledge of STEM lectures.

\section{Findings from Design Workshop}
We identified the two most commonly mentioned issues by participants and formulated five design recommendations for creating high-quality vicarious dialogues. Additionally, we propose how LLM can be integrated into the educational dialogue authoring process.

\subsection{Challenges in Converting Video Lectures to Dialogue}
Two challenges were observed when instructors converted video lectures to dialogue.

\textbf{\textit{Challenge 1: Designing the overall structure of dialogues.}} We observed that the participants faced difficulties in designing the overall structure of the dialogue when creating from scratch. Participants mostly first struggled with which part of the lecture should be converted to dialogue. P3 mentioned that it was \textit{``difficult to figure out which parts of a monologue should be transformed into direct learner's questions"} and  P4 said it was \textit{``hard to decide when and how much dialogue to create"}. It poses the cold start problem when designing dialogues by considering the improvement of the vicarious learner's learning. 
Furthermore, participants struggled to determine the appropriate format for the dialogue as they were unsure how the dialogue format would affect learning outcomes.
P5 said that \textit{``while it was easy to convert the lecture into a simple question-and-answer format, I’m not sure if these would be meaningful dialogues for vicarious learners"}. P15 also mentioned, \textit{``If it ends up looking too similar to the original lecture format, converting the material to a dialogue format might not be necessary"}, asserting the need to define what kind of dialogue format would be helpful for vicarious learners in an online learning environment.

\textbf{\textit{Challenge 2: Anticipating direct learner’s utterances based on their level of understanding.}} Both instructors and experts needed help with designing a direct learner's utterances. This is evident from comments: \textit{``It is hard to add direct learners' misconceptions to dialogues effectively"} (P15) and \textit{``It was difficult to consider individual responses of the direct learners"} (P7).

\subsection{Design recommendations that should be considered while designing dialogue for vicarious learners.}
Based on the challenges above, we propose five dialogue design recommendations. 
Furthermore, we suggest four teaching strategies (Table~\ref{tab:teaching-strategies} in Appendix) validated by workshop participants as likely effective even in a video-based learning context among pre-defined guidelines based on literature (Section 3.1.2).

\textbf{DR1. Dynamic: Include various interaction patterns to reflect the dialogic dynamics between the tutor and tutee.} A vicarious dialogue should be structured with fast turn-taking and various utterance patterns (Table~\ref{tab:tutor_category}, Table~\ref{tab:tutee_category}) that capture the dynamism of an actual tutoring scenario. Moreover, P14 mentioned that \textit{"fast turn-taking is required to hold the attention of vicarious learners in online education, as it is more difficult to retain focus on digital learning platforms than in physical classrooms"}. Furthermore, instructors and experts often divided the tutor's lengthy utterances into smaller sub-dialogues between the tutor and the direct learner, highlighting the quick turn-taking in vicarious dialogues. 

\textbf{DR2. Academically productive: Encourage the metacognitive and constructive utterances of the direct learner to make a dialogue academically productive.} 
Direct learners' utterances should be pedagogically meaningful to enhance vicarious learners' learning and engagement. Most workshop participants consistently emphasized the influence of direct learners on vicarious learners throughout the dialogue design process. Notably, they stressed the importance of direct learners displaying "interactive engagement" in dialogues, as vicarious learners are highly likely to empathize with the direct learner's learning process. The term "interactive engagement" refers to the active engagement of direct learners both cognitively and metacognitively.

\textit{Direct learner's cognitive engagement:} 
P15 highlighted the importance of a tutor in a vicarious dialogue who should encourage active engagement by facilitating connections between direct learners' existing knowledge and the new material, citing \textit{Ausubel's meaningful learning theory}~\cite{ivie1998ausubel}.
In addition, P14 mentioned that \textit{``When the instructor links the learning contents with the learner's personal experiences, the transfer learning occurs more easily"}.

\textit{Direct learner's metacognitive engagement:} P15 and P9 proposed incorporating self-assessment and explanations of understanding from the direct learner into vicarious dialogues:  \textit{``When a direct learner self-assesses their level of understanding or performs self-summarization, a vicarious learner could potentially check their comprehension"}. In addition, P15 suggested that a tutor continuously promotes the direct learner's metacognition. This guide aligns with the findings that in an ITS ~\cite{anderson1985intelligent,ma2014intelligent}, the constructive actions of a direct learner, such as answering based on what they learned from the instructor's scaffolding and asking deep-level reasoning questions ~\cite{ma2014intelligent,kulik2016effectiveness}, significantly influenced the learning outcomes and participation of vicarious learners.

\textbf{DR3. Cognitively adaptive: Adapt the teaching strategies to the level of understanding of the vicarious learner, learning objectives, and lecture contents.} Previous literature suggests that strategies requiring higher cognitive engagement, like inducing cognitive conflicts and posing deep-level reasoning questions, benefit vicarious learners ~\cite{driscoll2003vicarious,craig2006deep,gholson2009exploring,craig2012promoting}. However, applying cognitively demanding strategies, like \textit{cognitive conflict} in Table~\ref{tab:teaching-strategies} in Appendix, may not always suit all learning materials or learners when converting lecture videos into dialogues. P15 noted that the choice of cognitive strategy may vary depending on the granularity of the learning content being transformed into a dialogue. In addition, he emphasized the importance of aligning cognitive strategies with learning objectives and the level of vicarious learners, stating that \textit{``Frequent placement of lighter, easily answerable questions and minimal use of cognitive strategies on important content could lower the cognitive load on vicarious learners"}.

\textbf{DR4. Purposeful: Define a learning objective for the vicarious learner and ensure that the learning objective is achieved through that dialogue.}
To create meaningful dialogue for vicarious learners, we recommend aligning the dialogue's goal with the vicarious learner's learning objective and illustrating the achievement of this objective through interactions between a direct learner and a tutor. P15 and P8 emphasized the importance of defining clear learning objectives for vicarious learners as an initial step in dialogue creation. Additionally, P8 highlighted that learning objectives should be intimately tied to the difficulties vicarious learners face. 

\textbf{DR5. Immersive: Utilize realistic teaching scenarios and match the direct learner's cognitive level with the vicarious learner's level.} 
We suggest considering two factors that can immerse vicarious learners in their vicarious interaction. 
\begin{itemize}
    \item \textit{Incorporate common teaching scenarios}: Some participants suggested using real classroom scenarios for vicarious learner engagement. For example, P11 proposed scenarios in which the direct learner is given an incorrect problem and asked to explain what is wrong and a situation where another learner responds correctly to the tutor’s question when a student gives wrong answers. P14 also suggested a scenario where a tutor makes the direct learner apply what they have learned in different examples.  
    \item \textit{Match cognitive levels}: Instructors and experts highlighted aligning the cognitive levels of direct and vicarious learners in lecture videos to benefit the vicarious learners.--- \textit{``Vicarious learners often lose interest when confronted with familiar material but are more likely to engage when unfamiliar or essential information is presented.''} (P12). Therefore, addressing vicarious learners' unfamiliar or challenging parts through direct learners’ dialogue could be an effective way to design meaningful and high-quality dialogue.
\end{itemize}

\subsection{Enhancing the Educational Dialogue Design Process with LLMs}
After establishing guidelines, we explored how instructors and experts used LLM-generated dialogues and developed evaluation criteria (Table~\ref{tab:eval_metric}) for evaluating their pedagogical quality based on how workshop participants assess the dialogues (Table~\ref{tab:eval_metric_workshop}). We also explored strategies for integrating LLMs into the educational dialogue design process.

\begin{table*}[]
\resizebox{0.88\textwidth}{!}{%
\begin{tabular}{l|l}
\hline
\textbf{Criteria}               & \textbf{Key questions}                                                                                                                                                                                                                                                   \\ \hline
\textbf{Dynamic}                & \begin{tabular}[c]{@{}l@{}}Are various interaction patterns (Table ~\ref{tab:tutor_category}, Table ~\ref{tab:tutee_category}) incorporated to reflect the dynamics of \\ real classroom dialogue?\end{tabular}                                                                                                          \\ \hline
\textbf{Academic Productivity}  & \begin{tabular}[c]{@{}l@{}}Is the teacher effectively eliciting the learner's metacognitive and constructive utterances \\ to ensure the discourse is academically \\ productive?\end{tabular}                                                                  \\ \hline
\textbf{Cognitive Adaptability} & \begin{tabular}[c]{@{}l@{}}Are the cognitive strategies used in the dialogue adaptively applied based on the vicarious \\ learner's level, learning objectives, and the lecture contents?\end{tabular}                                                                   \\ \hline
\textbf{Purposefulness}         & \begin{tabular}[c]{@{}l@{}}Is the learning objective of vicarious learners achieved through the dialogue between the \\ direct learner and teacher?\end{tabular}                                                                                                         \\ \hline
\textbf{Immersion}              & \begin{tabular}[c]{@{}l@{}}Does the dialogue represent realistic teaching scenarios and establish a direct learner's level \\ comparable to that of a vicarious learner, thereby improving vicarious learner engagement?\end{tabular}                           \\ \hline
\textbf{Usefulness}             & \begin{tabular}[c]{@{}l@{}}Is the dialogue satisfactory and useful, considering personal experience with students, what \\ an instructor wants to emphasize, and the instructor's usage context, such as the level of \\ vicarious learners being targeted?\end{tabular} \\ \hline
\textbf{Correctness}            & Are domain-specific words used accurately, and is the conversation content based on facts?                                                                                                                                                                               \\ \hline
\end{tabular}%
}
\Description{This table presents criteria for evaluating the pedagogical quality of LLM-generated dialogues. The table includes eight criteria: 'Dynamic,' 'Academic Productivity,' 'Cognitive Adaptability,' 'Purposefulness,' 'Immersion,' 'Usefulness,' 'Correctness,' and 'Key questions.' Criteria in bold were only used in building the \sysname{} system as described in Section 5. Each criterion is accompanied by key questions for evaluation. The table is labeled 'Table 5: Criteria when evaluating the pedagogical quality of LLM-generated dialogues.
}
\caption{Criteria when instructors evaluated the pedagogical quality of LLM-generated dialogues in our design workshop.}
\label{tab:eval_metric_workshop}
\end{table*}

\subsubsection{Utilization of LLM-Generated Dialogues}
We propose two ways in which the LLM could enhance the dialogue design process for vicarious learners. 
Firstly, it can provide pre-generated dialogues, stimulating instructors' ideation. P2 commented that using the LLM felt like it provided helpful guidelines, making it more effective than starting from scratch. Secondly, it can assist in modifying dialogues at different levels, refining sub-dialogues and crafting direct learners' responses. 
Participants proposed presenting expected responses at different levels (P12) and automating the process of generating questions from the direct learner's perspective (P2). 

Despite the LLM's advantages, the dialogue authoring process still requires active instructor involvement. In our observation, we have noted that instructors have their own set of criteria when designing high-quality dialogues. These criteria are based on their teaching experiences and can vary depending on the instructor's emphasis on specific aspects where they believe vicarious learners may face challenges. Guided by these personalized criteria, instructors designed and revised their dialogues.

Some instructors found the generated dialogues satisfactory because they aligned with their intended teaching points or teaching style. P13 chose the dialogue, stating \textit{``When teaching math, using fewer variables is better. So, I initially emphasized reducing the number of characters and utilizing known information. The dialogue aligns well with my problem-solving approach that focuses on minimizing variables"}. 
Some instructors didn't use the dialogues because the content didn't meet their quality criteria. 
For example, P11 made revisions to emphasize a specific point, stating, \textit{``The tutee's question: 'So, is x-2 the square root of 6?' is crucial in the problem-solving process. It would be helpful if the tutor followed up with a question like, 'What is the number that becomes 6 when squared?' to elaborate on this point"}.

\subsubsection{Criteria for Evaluating the Educational Dialogues}
Instructors evaluated the quality of LLM-generated dialogue based on seven criteria (Table~\ref{tab:eval_metric_workshop}). 
Five of these criteria aligned with the key factors to consider when designing educational dialogues (Section 4.2), while the other two criteria, \textit{Usefulness} and \textit{Correctness}, pertain to evaluating dialogues generated by the LLM. 


\subsection{Design Goals}
Based on LLM’s strengths and limitations in designing educational dialogue and criteria that instructors emphasized the most when evaluating the quality of dialogues (Table~\ref{tab:eval_metric_workshop}), we propose four design goals (DG):

\textbf{DG1.} Enable instructors to easily simulate direct learners easily.

\textbf{DG2.} Assist instructors in designing dialogues by referencing utterances generated at various levels of granularity. 

\textbf{DG3.} Assist instructors in creating dialogues that reflect the user's dialogue usage context and personal experience with students.

\textbf{DG4.} Ensure that instructors consistently monitor important considerations when designing vicarious dialogues.

\section{\sysname{}: A SYSTEM FOR AUTHORING VICARIOUS DIALOGUES FROM MONOLOGUE-STYLED LECTURE VIDEOS WITH LLM ASSISTANCE}

Based on our design goals from the workshop, we developed \sysname{}, an LLM-based system to assist instructors in crafting vicarious dialogues from their monologue-styled lecture videos. 
While LLM holds potential benefits for the dialogue design process, as detailed in Section 4.3.1, they may not be practically utilized in real educational settings if \textit{Correctness} and \textit{Usefulness} (Table~\ref{tab:eval_metric_workshop}) are not ensured. Thus, \sysname{} provides a collaborative authoring process between LLM and instructors, facilitating the generation of high-quality and correct vicarious dialogues. Based on our four design goals and observed dialogue design process in the workshop, this collaborative authoring process consists of three stages: (1) \textit{Initial Generation}, (2) \textit{Comparison and Selection}, and (3) \textit{Refinement}.


\begin{figure*}[!h]
    \centering
    \includegraphics[width=\textwidth]{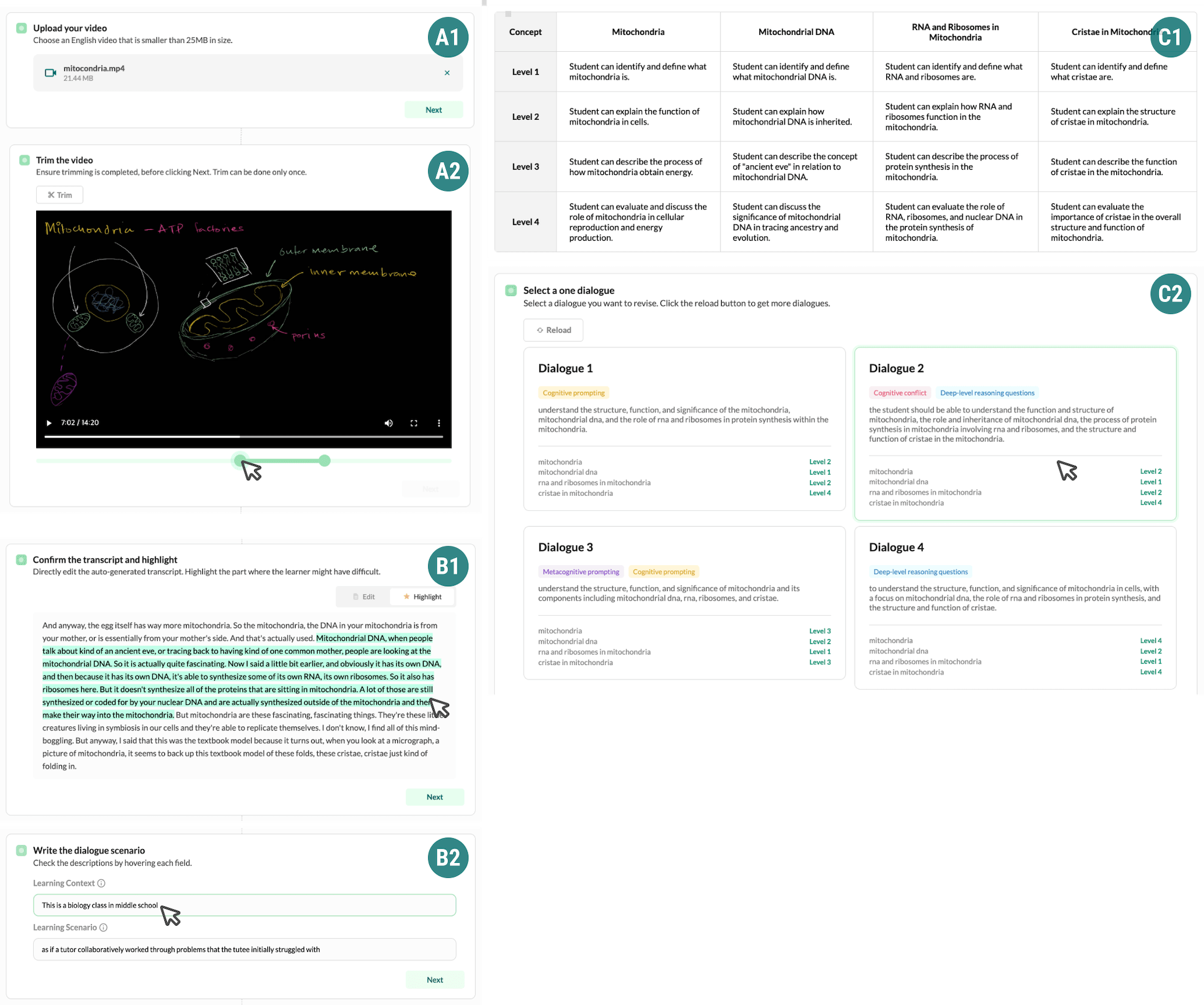}
    \caption{\sysname{}'s key components of \textit{Initial Generation} : (A1) User uploads lecture video; (A2) User trims a video section to convert ; (B1) User uses the \textit{highlighting} feature by selecting a part of the video transcript, where vicarious learners may face difficulty understanding ; (B2) User writes down the learning context and the scenario of dialogue that they want to depict in final dialogue, and \textit{Comparison and Selection} phase : (C1) \sysname{} shows a rubric table of learners' \textit{understanding level} regarding key concepts stated in the transcript; (C2) \sysname{} presents generated dialogues in the form of \textit{dialogue cards} comprising of core information from each dialogue.}
    \Description{Each components illustrate key feature of \sysname{}. After the user selects a section of the input video, \sysname{} shows up the auto-generated transcript, and (B1) the user highlights a part of the transcript where the learner might have difficulty understanding. (B2) Then user inputs desired dialogue context and scenario. Based on user's inputs, \sysname{} generates four dialogues in a form of (C2) dialogue cards with (C1) knowledge rubric table. Rubric table depicts learner's understanding level of key concepts in 4-level scale. Each learners in dialogue has one level state for each key concepts and it accounted on dialogue cards.}
\label{fig:system_interface_1}
\end{figure*}

To motivate \sysname{}’s design, we describe a usage scenario where an instructor collaborates with LLM to author dialog through \sysname{}.
A high school biology teacher, Sophia requires her students to watch recorded lectures before class. Sophia wants to make sure that students easily understand parts of the lectures with the most common misconceptions. In this context, she uses \sysname{} to transform the sections in her recorded lecture where misconceptions frequently occur into dialogues so that her students gain a better understanding. Thus, she uploads her lecture video to \sysname{} \textbf{(A1, Figure~\ref{fig:system_interface_1})} and selects the sections she wants to transform into dialogues \textbf{(A2)}.

\textbf{Initial Generation}. She then highlights areas where her students might develop misconceptions or key examples she wants to emphasize in the dialogue \textbf{(B1, Figure~\ref{fig:system_interface_1})}. Sophia aims to design the dialogue scenario as if it is occurring in a high school biology class, where a teacher addresses the direct learner's misconceptions in the dialogue \textbf{(B2)}. Upon highlighting, \sysname{} generates four dialogues reflecting the dialogue scenario.  

\textbf{Comparison and Selection.} \sysname{} shows generated dialogues with an `understanding level rubric'~\textbf{(C1, Figure~\ref{fig:system_interface_1})} that shows four levels of learners' understanding for each key concept in the selected part and the `dialogue cards~\textbf{(C2)}' that contains key information of each dialogue. Sophia compares each dialogue, considering the knowledge levels of the direct learner for each concept illustrated in the dialogue cards~\textbf{(C2)}. She then chooses to modify `Dialogue 2' because it highlights the misconceptions she wants to include. 

\begin{figure*}[!h]
    \centering
    \includegraphics[width=\textwidth]{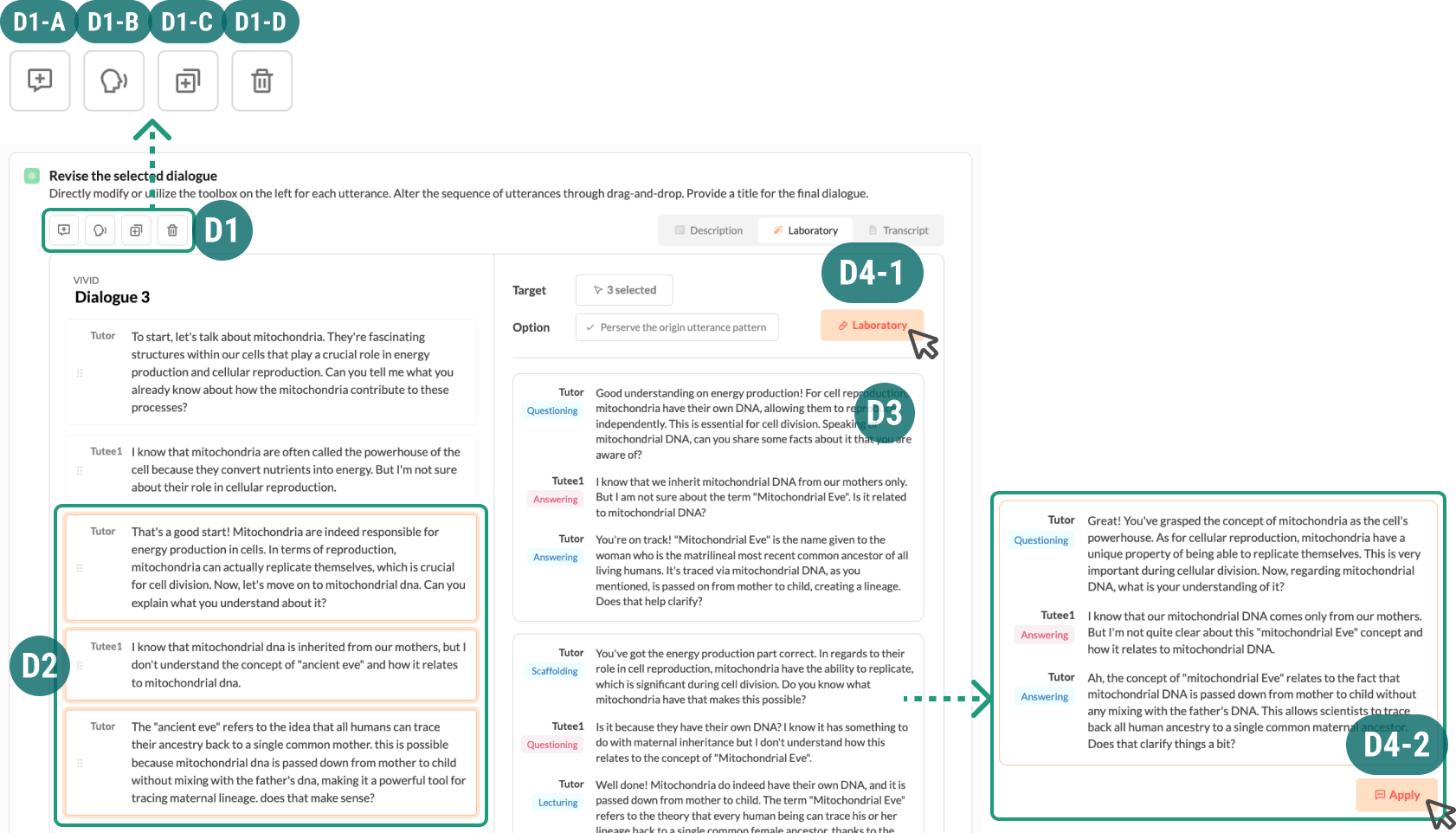}
    \Description{A screenshot illustrating key components of the 'Refinement' phase in \sysname{}. (D1) Users can edit utterance content directly or through basic editing tools. (D2) Users can utilize the 'laboratory' feature by selecting consecutive utterances and clicking the 'laboratory' button. (D3) \sysname{} generates four variations of sub-dialogues as suggestions. (D4-2) Users can replace the original utterances with a variation by clicking a button.}
    \caption{\sysname{}'s key components of \textit{Refinement} phase : (D1) User can edit each utterance content directly or using basic editing tools; (D2) User can use \textit{laboratory} feature by selecting consecutive utterances and clicking (D4-1) \textit{laboratory} button ; (D3) \sysname{} suggests four variations of sub-dialogues as a result; (D4-2) \textit{apply} button ; User can replace the original utterances with a variation by clicking button.}
\label{fig:system_interface_2}
\end{figure*}

\textbf{Refinement.} Sophia modifies `Dialogue 2' by adding questions in the tutor's utterance to address direct learner's misconceptions. She clicks the \textit{Generate} button \textbf{(D1-A, Figure~\ref{fig:system_interface_2})} to add a new utterance. However, she is unsure what answers the direct learner could provide for these newly added questions. To view different examples of how the learner might respond, she first selects the learner's utterance that she wants to see more variations of clicked sub-dialogue \textbf{(D2)}. Afterward, she clicks the \textit{Laboratory} button \textbf{(D4-1)}, and \sysname{} generates four variations of the chosen utterances. 

After reviewing the results, she wants to replace the existing utterances with new ones that better represent the learner's misconceptions. She clicks the \textit{Apply} button \textbf{(D4-2)} to replace the previous utterances with new ones. This allows Sophia to create a dialogue where misconceptions are effectively addressed in the final dialogue.

\begin{figure*}[!h]
    \centering
    \includegraphics[width=\textwidth]{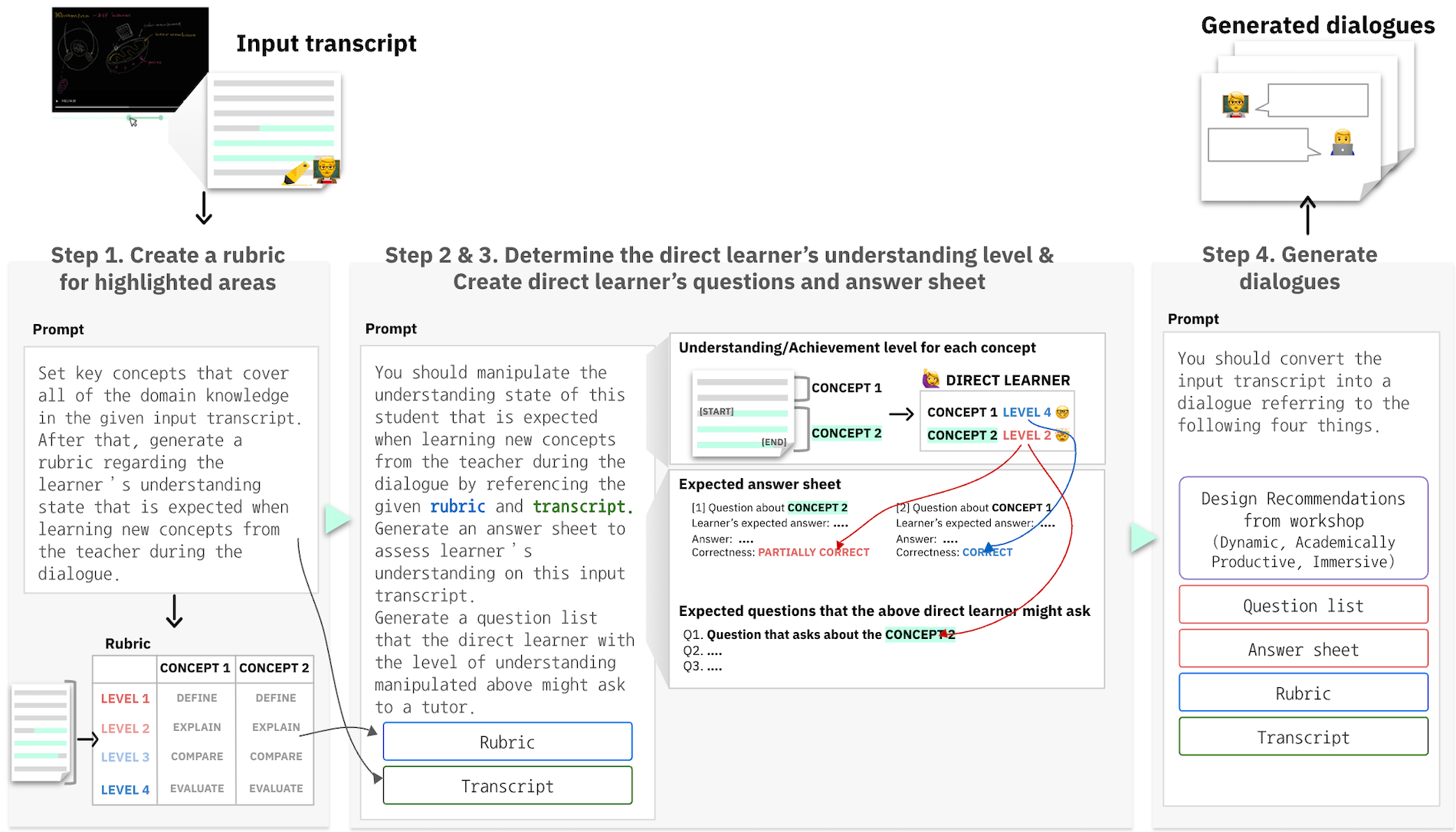}
    \caption{Overview of prompting pipeline for \textit{Initial Generation} phase. Each step corresponds to following subsections: (1) Create a rubric for highlighted areas, indicating the learner's understanding level for each concept ; (2) Determine the direct learner's understanding level using the highlighted parts and the rubric ; (3) Create an answer sheet consisting of the learner’s expected answers to the tutor’s questions and questions showing where the learner struggles ; (4) Generate dialogues based on the guidelines.}
    \Description{Each step indicates prompting strategies for the initial generation phase. (1) Based on the user's highlight and dialogue scenario inputs, the system generates a rubric of key knowledge and set a learner's understanding level using the rubric. This rubric also shows up to user in a table form. (2) To apply the highlight section to dialogue, a learner has a lower level for the corresponding concept to the highlight section, which means lower understanding. (3) After the learner's level is set, the system creates a question and answer sheet for the learner regarding key concepts. The result of this step is utilized in dialogue generation.}
\label{fig:prompting_strategy}
\end{figure*}

\subsection{Initial Generation}

VIVID initially creates various dialogues for instructors to choose the one that aligns best with their intention for converting monologue to dialogue as we found that the LLM-generated dialogues can be utilized as prototypes in the process of educational dialog design (Section 4.3.1). Notably, our LLM-based pipeline of the Initial Generation stage is designed to generate dialogues that satisfy the most emphasized characteristics by workshop participants, which are \textit{Dynamic, Academically Productive}, and \textit{Immersive} (DR1, DR2, and DR5 in Section 4.2). Furthermore, when generating dialogues, \sysname{} reflects instructors' needs in our pipeline, making instructors easily simulate direct learners with knowledge levels similar to their target vicarious learner (DG1 in Section 4.4). Thus, the Initial Generation stage consists of four steps to generate dialogues that finely adjust the direct learner's knowledge state based on the instructor's needs.
We determined our final prompts (further details are in the Supplemental Material) by evaluating the quality of various dialogues based on our evaluation criteria (Table~\ref{tab:eval_metric}).

\subsubsection{Step 1. Create a rubric for highlighted areas, indicating the learner's understanding level for each concept.}
DR5 (\textit{Immersive}) in Section 4.2 suggests that the dialogue should align the cognitive level of direct learners with vicarious learners. 
\textit{Highlighting} feature allows instructors to highlight sections in the script that vicarious learners might find challenging. It reflects the intention of instructors to convert the dialogue for a specific level of vicarious learners. Therefore, \sysname{} leverages the highlighted sections to make assumptions about the level of vicarious learners and uses it to model the direct learner (DR5 in Section 4.2).

Before configuring the direct learners' understanding state, we extract the core concepts of the selected area in the transcript and divide the direct learners' possible understanding state of each concept into four levels. 
These levels are based on the cognitive domain of Bloom's taxonomy ~\cite{forehand2010bloom} as it has been used by instructors to design, assess, and evaluate student's learning~\cite{lord2007moving}. \sysname{} then generates four \textit{understanding levels} for each key concept with LLM and presents them in a rubric format \textbf{(B1)} (Figure ~\ref{fig:system_interface_1}).
The \textit{understanding level} here refers to the understanding state expected of direct learners when they learn new concepts from the instructor during the dialogue. 

\subsubsection{Step 2. Determine the direct learner's understanding level using the highlighted parts and the rubric.}
The highlighted parts present the concepts that the direct learner may not fully comprehend after the tutor's explanation in the dialogue. We set the direct learner's understanding level based on the highlighted concepts, using `level 1', `level 2', or `level 3' in the generated rubric to indicate the direct learner's knowledge deficits. The direct learner is prompted at the highest understanding level, `level 4' for unhighlighted areas. 

\begin{figure*}[!h]
    \centering
    \includegraphics[width=0.97\textwidth]{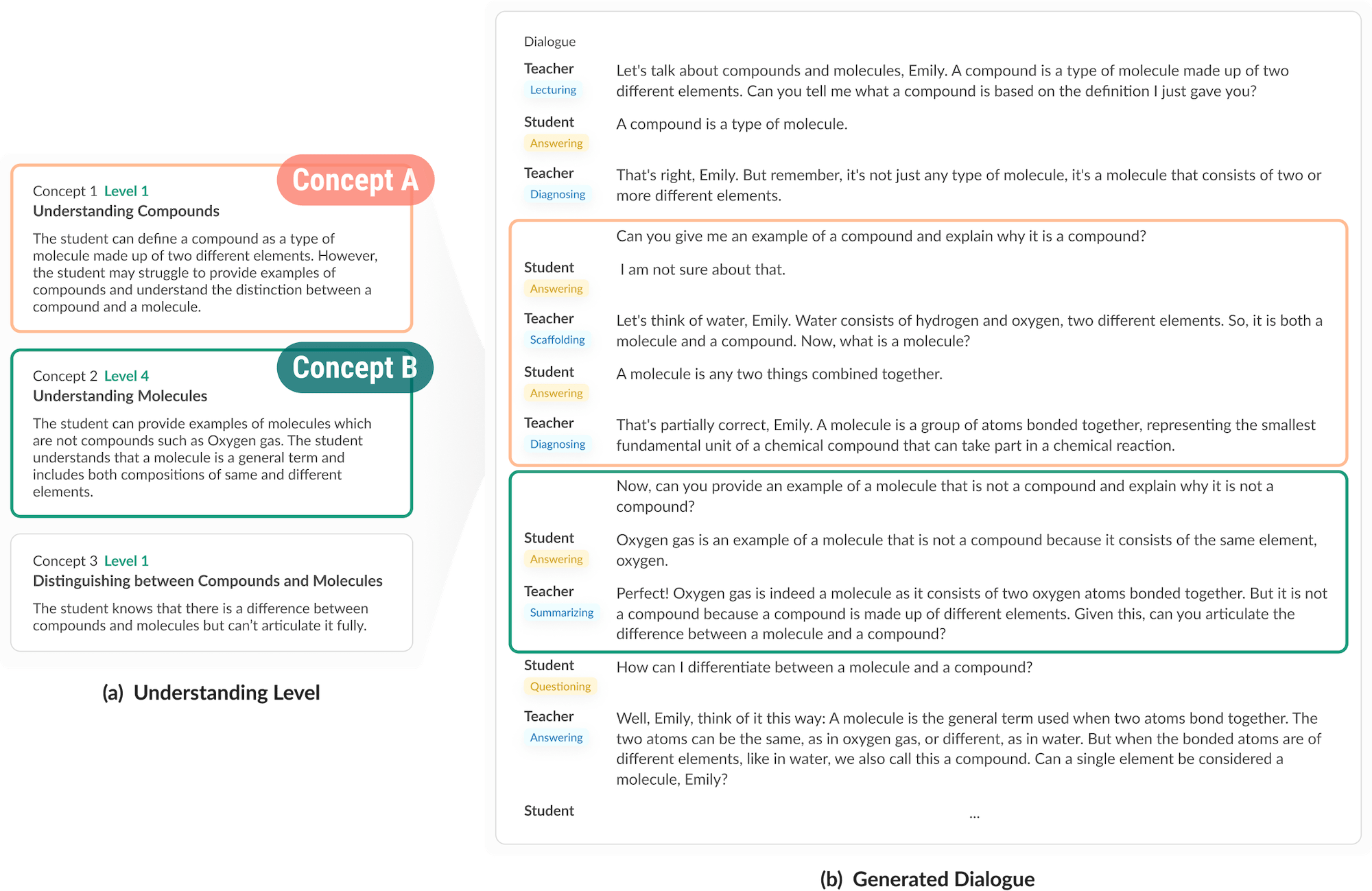}
    \caption{Example of generated dialogue regardless of the prerequisite relationships between key concepts. Concept A is a prerequisite for Concept B. During the conversation, the direct learner didn't understand the Concept A initially, but grasped it through question-and-answer, and answered Concept B correctly later.}
    \Description{Left is a direct learner's understanding level of key concepts in generated dialogue. Concept A is prerequisite knowledge of Concept B, and the learner has a higher understanding level regarding Concept B. Right shows how the learner's understanding level is applied to generated dialogue regardless of the concept's relationship. During the conversation, the student didn't understand concept A initially, but grasped it through question-and-answer, and answered concept B correctly.}
\label{fig:prerequisite_example}
\end{figure*}

The process of determining a direct learner's understanding level didn't consider prerequisite relationships between concepts to generate a dialogue that reflects varied levels of comprehension of each concept, as shown in Figure~\ref{fig:prerequisite_example}.
For example, consider a case where Concept A is a prerequisite for Concept B. Even if the LLM model sets Concept A at `level 1' and Concept B at `level 4', a scenario can be designed where the learner studies Concept A with the teacher to fill the knowledge gap (level 1) and then responds well to Concept B (level 4). 

\subsubsection{Step 3. Create an answer sheet consisting of the learner’s expected answers to the tutor’s questions and questions showing where the learner struggles.}
We designed our prompt to create expected questions and responses to the instructor's questions when the direct learner is in a specific knowledge deficit state. The expected answer sheet was designed in a descriptive format to reflect the learner's nuanced understanding. We prompted an LLM to manipulate the expected answers to the instructor's questions concerning the learner's knowledge level for each concept. We also designed a prompt to generate questions that direct learners might struggle with the concepts set to a low level.  

\subsubsection{Step 4. Generate dialogues.}
The final dialogues are generated through prompts based on the following three elements as shown in Figure~\ref{fig:prompting_strategy}: (1) Adjusted direct learner's knowledge state information through Step 1 to Step 3 to achieve \textbf{Immersive (DR5)}, (2) Key utterance categories of a tutor and a tutee in Table~\ref{tab:tutee_category} and Table~\ref{tab:tutor_category} to achieve \textbf{ Dynamic (DR1)}, and (3) Key teaching strategies described in Table~\ref{tab:teaching-strategies} to achieve \textbf{Academic Productive (DR2)}.

\begin{figure*}[!h]
    \centering
    \includegraphics[width=\textwidth]{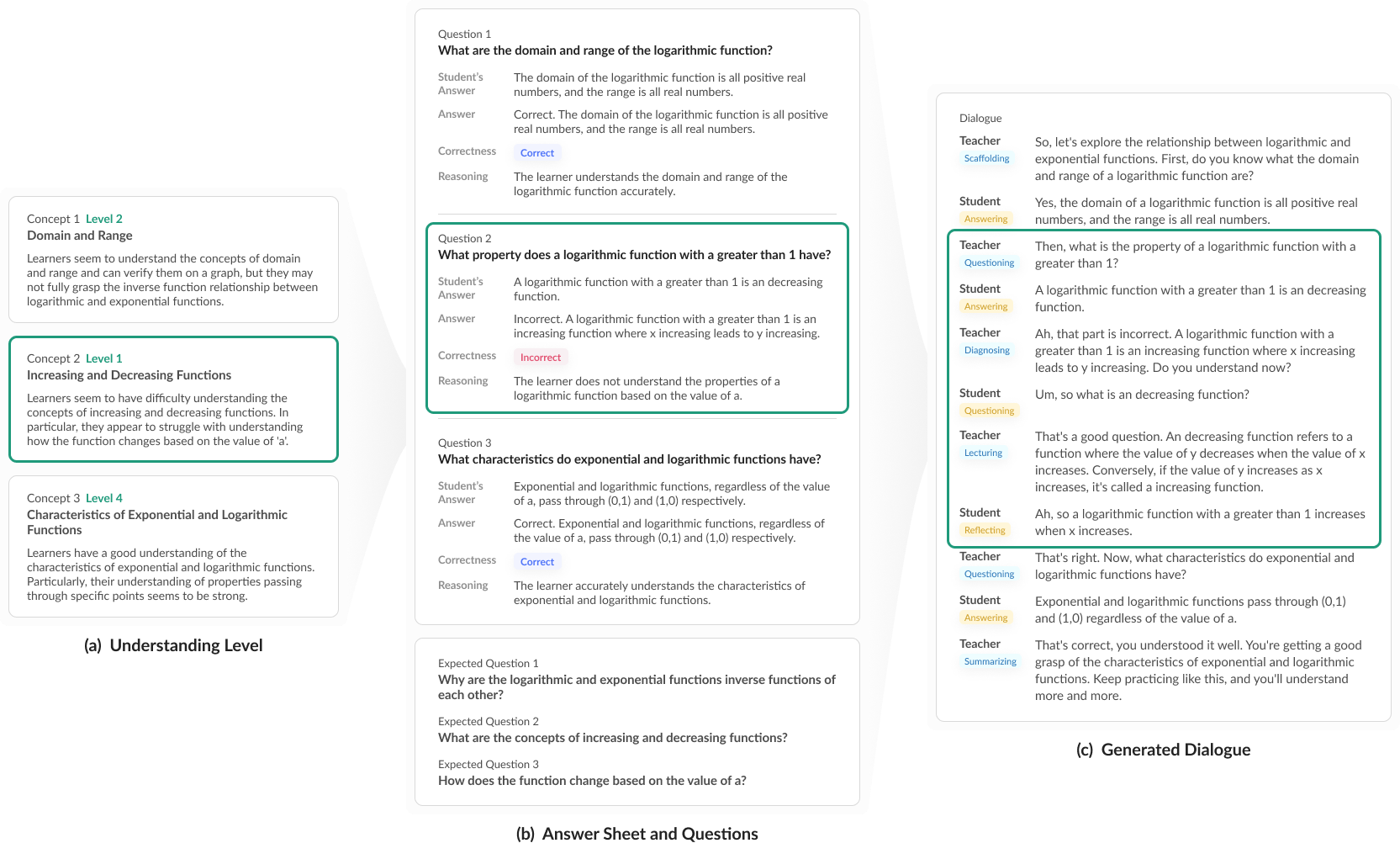}
    \caption{\textit{Initial Generation} pipeline. (a) \textbf{Understanding level}: Example of the direct learner's understanding level using the highlighted parts and the rubric, (b) \textbf{Answer Sheet and Questions}: Example of the answer sheet consisting of learner's expected answers to the tutor's questions and expected questions of direct learner, (3) \textbf{Generated Dialogue}: Example of final dialogue based our guideline-based prompt. The green box shows how the concept that set in \textit{level 1} reflects on the final dialogue.}
    \Description{This is an example of an initial generation pipeline. (a) shows how the learner's knowledge level is set regarding key concepts. (b) is an answer sheet of the learner, and (c) is a generated dialogue. Highlighted box shows how the pipeline works on certain concept in dialogue. As the learner has a low level on Concept 2, the answer sheet of the learner shows an incorrect answer for Concept 2. This deficit of knowledge is depicted well in the final dialogue result.}
\label{fig:initial_generation}
\end{figure*}

\subsection{Comparison and Selection}
In \textit{Comparison and Selection} stage, \sysname{} provides the instructors with an \textit{Understanding level rubric} \textbf{(B1)} and \textit{Dialogue cards} \textbf{(B2)} (Figure ~\ref{fig:system_interface_1}) to enable monitoring and selecting based on the criteria that were important during \textit{Initial Generation} stage (DG4 in Section 4.4). Each \textit{dialog card} \textbf{(B2)} contains the primary information of the dialogue, such as the direct learner’s understanding level of each concept, key teaching strategies, and key dialogue patterns. Besides, \textit{Understanding level rubric} represents a four-level understanding state for each key concept appearing in the selected part in the transcript.

\subsection{Refinement}
\subsubsection{Basic tools for instructor's direct refinement.}
In the workshop, we observed that instructors were proficient in using existing dialogue content, like breaking down lengthy tutor utterances into smaller segments or incorporating script contents into dialogue. To facilitate this kind of authoring, \sysname{} provides four basic functions: \textit{add} \textbf{(D1-a)}, \textit{duplicate} \textbf{(D1-b)}, \textit{delete utterance} \textbf{(D1-c)}, and \textit{change speaker} \textbf{(D1-d)}. As visible in \textbf{(D1)}, each utterance box in the final dialogue is clickable and can be moved with drag-and-drop (Figure ~\ref{fig:system_interface_2}). Additionally, we aimed to enhance the \textit{Correctness} of the dialogue through direct refinement.

\subsubsection{LLM-based refinement tool: \textit{Laboratory} }
In addition to basic functions, \sysname{} offers the \textit{Laboratory} tool \textbf{(D4-1)} that provides alternatives \textbf{(D3)} for the selected sub-dialogues \textbf{(D2)} through LLM (Figure ~\ref{fig:system_interface_2}). It is designed to address the instructor's challenges in developing direct learners' utterances while considering their understanding level (\textit{Challenge 2} in Section 4.1) and achieve DG3 (Section 4.4).
To do this, we designed the prompt used in \textit{Laboratory} tool while maintaining four key elements except for the original dialog patterns (in Supplemental Material): (1) learner's level of the selected dialogue in the \textit{Comparison and Selection} phase, (2) dialogue context, (3) main learning contents, and (4) the number of turns. On the other hand, we diversified the dialogue patterns, reflecting utterance categories in Table~\ref{tab:tutee_category} and Table~\ref{tab:tutor_category} in our prompt. When the instructor clicks the \textit{Apply} button \textbf{(D4-2)}, the selected sub-dialogue \textbf{(D2)} is replaced with the new sub-dialog \textbf{(D4-2)}. 

\subsection{Implementation}
\sysname{} is implemented using React~\footnote{\url{https://react.dev/}}, connected to a Flask~\footnote{\url{https://flask.palletsprojects.com/}}-based back-end server that utilizes GPT API. Whisper~\cite{radford2023robust}, an automatic speech recognition model by OpenAI, auto-generated the script of the section that the instructor chose from the lecture video (\textbf{B1} in Figure~\ref{fig:system_interface_1}). To address limitations in text-to-speech (TTS) models like noise or language and get more precise dialogue conversion, \sysname{} allows instructors to modify the TTS output directly during the \textit{Initial Generation} stage. 

Subsequently, the system harnessed the API of the latest trained GPT-4, OpenAI's advanced language model, to generate the rubric, learner's knowledge level, predicted answer sheet, and the final dialogue. Considering the importance of model accuracy in an educational context, we conducted prompt engineering experiments using GPT-3.5 and GPT-4. We chose to use GPT-4 due to its superior generation quality. We set a temperature of 0.65 for the rubric generation, which was empirically determined through trial to maintain consistency, and used the default temperature for other features.

\section{Evaluation}
To evaluate the performance of \sysname{} in designing high-quality educational dialogues, we conducted a two-fold evaluation --- user study and technical evaluation. In this section, we provide the details of each evaluation and results, respectively.

\begin{table*}[]
\resizebox{0.5\textwidth}{!}{%
\begin{tabular}{l|l|l|l|l}
\hline
ID  & Gender & Age & Career                      & Subject taught  \\ \hline
P1  & F      & 50s & 30 years                    & Math            \\
P2  & M      & 20s & 2 years                     & Science         \\
P3  & F      & 20s & 1 year                      & Science         \\
P4  & M      & 40s & 15 years                    & Math            \\
P5  & M      & 30s & 7 years                     & Engineering     \\
P6  & M      & 20s & 2 years                     & Math            \\
P7  & M      & 20s & 4 years                     & Engineering     \\
P8  & M      & 20s & 5 years                     & Math            \\
P9  & F      & 20s & 4 years                     & Math            \\
P10 & F      & 20s & 2 years                     & Science         \\
P11 & M      & 20s & Graduated teacher's college & Math            \\
P12 & F      & 20s & 1 year                      & Math \& Science \\ \hline
\end{tabular}%
}
\Description{This table displays the demographics, career information, and subjects taught by participants in the User Study. The table includes five columns: 'ID,' 'Gender,' 'Age,' 'Career,' and 'Subject taught.' The 'ID' column lists participant IDs, 'Gender' indicates gender (F for female, M for male), 'Age' specifies the age group, 'Career' describes the years of career experience, and 'Subject taught' denotes the subjects taught in the classroom. The table is titled 'Table 6: User Study participants' demographics, career, and their subjects taught in the classroom.
}
\caption{User study participants' demographic, career, and their subject taught in the classroom.}
\label{tab:user_study_p}
\end{table*}

\subsection{User Study}
\sysname{} is designed to autonomously generate \textit{Dynamic, Academically Productive,} and \textit{Immersive} dialogues between a tutor and a direct learner and support instructors in efficiently modifying them. To validate the efficacy of \sysname{}, we conducted a within-subjects experiment with 12 participants, comparing it with the baseline system that lacks \sysname{}'s core features. 

\subsubsection{\textit{\textbf{Study Setup}}} 
Participants were asked to transform a part of the lecture video chosen by the authors, into a dialogue using the systems under each condition. Participants experienced both conditions with different videos in a counterbalanced order to prevent bias and ensure validity. We analyzed user behavior logs, post-survey, and interview data to understand how our system supported the authoring process. 

\textit{\textbf{Baseline Condition}} 
The following text describes how the Baseline system differs from the \sysname{} system regarding the four design goals. In the \textit{Initial Generation} phase of the Baseline, it utilized a simple prompt (the detailed prompt is in the Supplemental) to create a dialogue that did not reflect the learner's understanding. Thus, the entire process of adjusting direct learner's knowledge through \textit{Highlighting} feature (in Section 5.1.1) was excluded. During the \textit{Compare and Selection} stage of the Baseline, the \textit{summarized card function} and \textit{understanding level rubric} were excluded from \sysname{}, enabling compare and selection of one out of four dialogues for revision without any background information about the generated dialogues. In the \textit{Refinement} phase of the Baseline, the \textit{laboratory function}, which offers multiple contextual alternatives for the sub-dialogue selected by the instructor, was removed.

\textit{\textbf{Lecture Selection}} 
The clarity of the lecture video can have an impact on the quality of the resulting dialogue. Other factors, such as the length of the video, the difficulty of the content, and the subject matter, can also influence the dialogue creation process. Therefore, when selecting lecture videos, we carefully considered the lecturer's explanation style and balanced the educational content and level of difficulty across all conditions. All videos were aimed at secondary school students, and we chose lecture content with similar prerequisite levels and granularity. Each video was in Korean and was approximately 10 minutes in length.

As we targeted STEM subjects, we selected two science and two mathematics lectures to use: topics for the science lecture were \textit{generation of waves} ~\footnote{\url{https://www.youtube.com/watch?v=u0KO1rm8neI}} and the \textit{refraction of waves} ~\footnote{\url{https://www.youtube.com/watch?v=64dZGBCELBc}}, and topics for mathematics were \textit{exponential function} ~\footnote{\url{https://www.youtube.com/watch?v=FBAgxbQ931Y}} and \textit{logarithmic function} ~\footnote{\url{https://www.youtube.com/watch?v=I_H04p9HHcI}}.
Mathematics lectures are presented in the format of writing board screencasts with voice-over~\cite{chorianopoulos-2018}. Science lectures are presented in the same format but based on slides. Each video follows a monologue-style lecture, where the instructor teaches without direct learners. The audio recording quality of all videos is at a level where the instructor can watch the lectures without any issues.

\textit{\textbf{Participants}} We recruited 12 participants via social media platforms, including the local community for instructors. The participants were required to 1) teach STEM subjects, 2) have experience in designing online lectures or using them in their classes, and 3) be either school teachers or part-time instructors. We recruited participants for \sysname{} without considering teachers' experience levels, as \sysname{} is designed to support teachers regardless of their experience. All sessions were carried out via Zoom, and participants were compensated at a rate of 45,000 won per hour (equivalent to 34 USD).

\subsubsection{\textit{\textbf{Study Procedure}}} 
The study consisted of three tasks, followed by a post-task survey and interview.

\textit{Task 1. Eliciting ambiguous intent for the direct learner design.} 
The participants were asked to convert a challenging section of a lecture into a dialogue that would help vicarious learners better understand the topic. In \sysname{} condition, the instructors had to select the specific contents that might be difficult for vicarious learners and convert them into dialogue using the \textit{highlighting} feature.
The specific guidelines on how the \textit{highlighting} feature would affect the dialogue generation pipeline were not provided. On the other hand, instructors were only asked to choose where to convert without the \textit{highlighting} feature in the Baseline condition. They then wrote about the teaching scenarios they wanted to depict in a dialogue.

\textit{Task 2. Comparing and selecting a dialogue to revise.}  Participants in the \sysname{} condition referred to dialogue cards and rubric to select one dialogue from four generated in \textit{Initial generation} stage for revision. However, in the Baseline condition, instructors had to choose a dialogue that was designed without considering the direct learner, and they could not consider \textit{rubrics} and information regarding the direct learner in choosing a dialogue.  

\textit{Task 3. Revising a chosen dialogue.} Participants in both conditions could refine the selected dialogue, employing the system's basic refinement functions. In the \sysname{} condition, participants could use the \textit{laboratory} feature (Section 5.3.2) to refine their dialogue.

\begin{figure*}[!h] 
\centering 
\includegraphics[width=\textwidth]{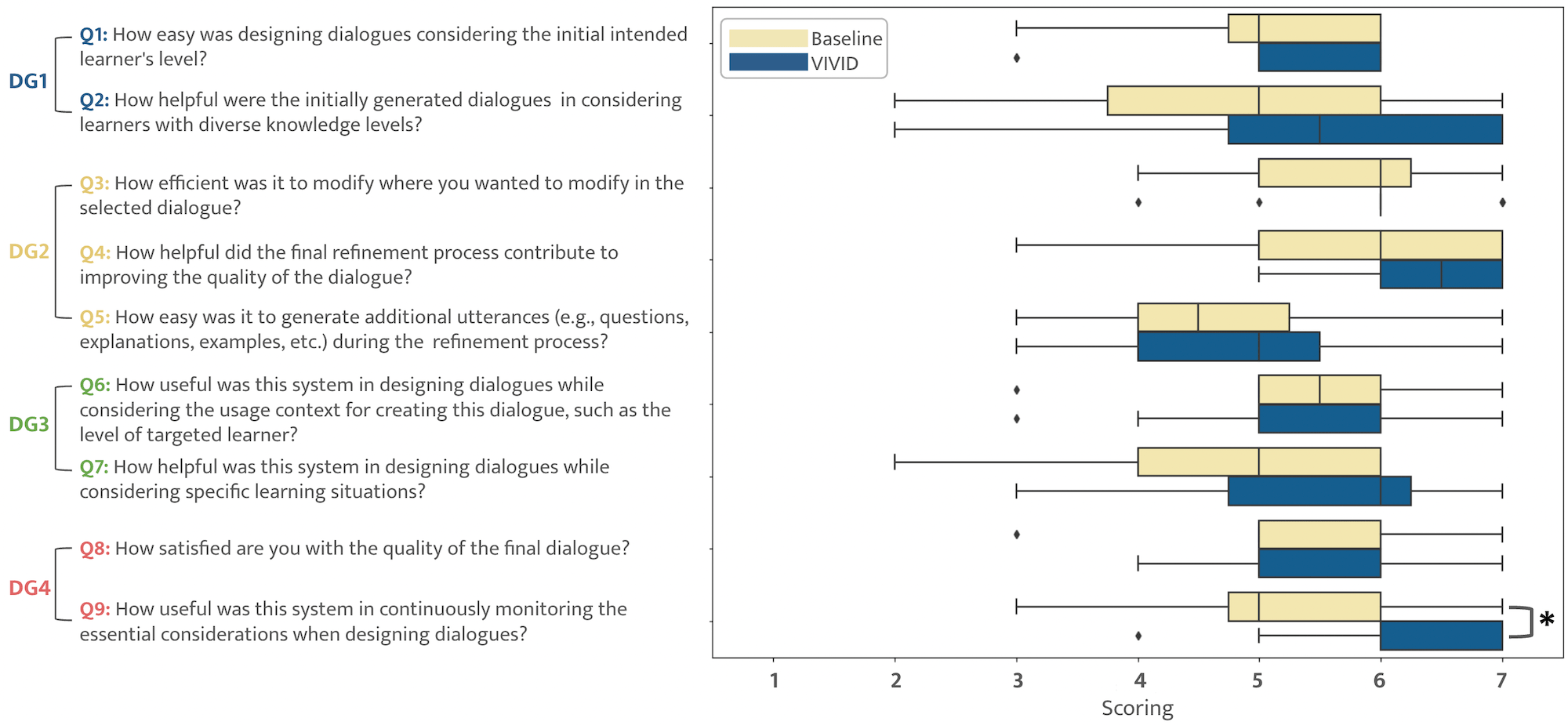} 
\caption{Post-task survey results on nine questions regarding task experiences. Each question was evaluated on a 7-point Likert scale. Treatment is corresponding to the \sysname{}.}
\label{fig:post_survey}
\end{figure*}

\textit{Post-task survey and interview} After completing the tasks with both conditions, participants were asked to fill out a 7-point Likert Scale questionnaire that consists of nine questions to evaluate whether each feature of the system under each condition well reflected the design goals in Section 4.4 for creating quality educational dialogue and whether it produced quality dialogue (Figure~\ref{fig:post_survey}). We conducted a semi-structured interview to understand participants' experiences with each system, the generated conversation, and the dialogue authoring experiences.

\subsection{\textit{\textbf{User Study Results}}} 

Despite the overall high utility of the Baseline (Figure~\ref{fig:post_survey}), nine out of 12 participants found VIVID to be better for designing vicarious dialogues due to its unique features such as \textit{rubric}, \textit{dialogue card}, and \textit{laboratory} features. Notably, instructors considered VIVID to be significantly more helpful than the Baseline in monitoring important factors in dialogue design, as shown in Q9 of Figure~\ref{fig:post_survey}. However, apart from this, no other significant differences in usefulness were observed. 

\subsubsection{\sysname{} helped participants monitor essential considerations when designing conversations.}
Participants rated \sysname{} (M = 6.1, SD = 0.9) as significantly more useful in assisting them in monitoring key considerations persistently in dialogue design (Q9 in Figure~\ref{fig:post_survey}) compared to the Baseline (M = 5.2, SD = 1.3, p = 0.04, Wilcoxon signed-rank test). Furthermore, while instructors felt that \sysname{} (M = 5.5, SD = 1.31) was more useful than the Baseline (M = 4.75, SD = 1.13) in considering specific teaching scenarios when designing dialogues (Q7), the difference was not statistically significant (p = 0.07, Wilcoxon signed-rank test). In terms of satisfaction with dialogue quality (Q8), there was a minimal difference between \sysname{} (M = 5.7, SD = 0.94) and the Baseline (M = 5.6, SD = 0.95). Although \sysname{} played a significant role in managing the educational dialogue design process, both conditions resulted in similar satisfaction levels due to manual refinement.

\subsubsection{\sysname{} helped instructors simulate a direct learner with diverse levels of understanding.}
Although the difference in Q2 (Figure~\ref{fig:post_survey}), which evaluates how helpful the initially generated dialogue was in considering learners of various knowledge levels, was not significant, \sysname{} (M = 5.3, SD = 1.6) had a higher average than the Baseline (M = 4.6, SD = 1.56). In addition, some instructors highlighted \sysname{} was better at selecting a suitable dialogue by considering the direct learner's knowledge level for each dialogue than Baseline. P1 mentioned, \textit{``\sysname{} was more conducive to constructing a lesson script optimized for the target learner as it clearly indicates the learning stage compared to the Baseline.''}. Furthermore, P4 stated, \textit{``\sysname{} was preferable as it allows selection and refinement according to the learner’s level by showing rubric, so it was helpful for selecting dialogues with an appropriate difficulty level.''}. Notably, P5 and P11 mentioned that the understanding level rubric provided with the dialogue cards allowed them to consider the direct learner's level more specifically when choosing a dialogue.

\subsubsection{\sysname{}’s \textit{laboratory} feature helped instructors better predict the direct learner's responses and improve the dialogue's pedagogical quality.}
Eight of eleven instructors who used the \textit{laboratory} feature were satisfied with this feature. One instructor did not use this feature. 
Some instructors highlighted how this feature positively impacted the dialogue quality. We observed that the \textit{laboratory} feature helped instructors explore the design space of dialogues while considering possible responses from direct learners. P1 said, \textit{``Especially regarding the utterances of direct learners, it was difficult for the participants to imagine what questions the learner would ask, but through this feature, I was able to consider various learning situations and learner's responses that I hadn't thought of before.''}. P5 also mentioned, \textit{``I could consider answers and questions that direct learner might have from a wider range of perspectives''}.

\aptLtoX{\begin{table*}[]
\begin{tabular}{c|l|l}
\hline
\textbf{Criteria}   & \multicolumn{1}{c|}{\textbf{\begin{tabular}[c]{@{}c@{}}Statement  (7-point Likert Scale)\end{tabular}}}                                                                                                                                    & \multicolumn{1}{c}{\textbf{\begin{tabular}[c]{@{}c@{}}Measuring questions  (Pairwise Comparison)\end{tabular}}}                                                                                                                              \\ \hline
\textbf{Dynamic} & SD1. The dialogue demonstrates clear and fast turn-taking. & QD1. Which one demonstrates clearer and faster turn-taking?                                                                                                                                                                                    \\
& \begin{tabular}[c]{@{}l@{}}SD2. The dialogue utilizes diverse interaction patterns  between a tutor and tutees.\end{tabular}                                                                                                               & \begin{tabular}[c]{@{}l@{}}QD2. Which one utilizes more diverse interaction patterns  between a tutor and a tutee?\end{tabular}                                                                                                              \\ \hline
\textbf{\begin{tabular}[c]{@{}c@{}}Academic Productivity\end{tabular}}
& \begin{tabular}[c]{@{}l@{}}SAP1.  The dialogue encourages the learner's  cognitive engagement (e.g., asking about what they've  learned, asking various types of questions, and inquiring  about a student's experiences)\end{tabular} & \begin{tabular}[c]{@{}l@{}}QAP1. Which one encourages the learner's cognitive engagement more? (e.g., asking about what they've  learned, asking various types of questions, and inquiring  about a student's experiences)\end{tabular} \\
  & \begin{tabular}[c]{@{}l@{}}SAP2. The dialogue prompts a student's metacognitive  thinking.\end{tabular}                                                                                                                                    & \begin{tabular}[c]{@{}l@{}}QAP2. Which one prompts a learner's metacognitive  thinking more?\end{tabular}                                                                                                                                    \\ \hline
\textbf{Immersion} & \begin{tabular}[c]{@{}l@{}}SI1. The dialogue appears to describe a specific and  natural learning situation.\end{tabular}                                                                                                                  & \begin{tabular}[c]{@{}l@{}}QI1. Which one describes a more specific and natural  learning situation?\end{tabular}                                                                                                                            \\
  & \begin{tabular}[c]{@{}l@{}}SI2. The dialogue reveals and addresses a learner's  knowledge deficits more clearly.\end{tabular}                                                                                                              & \begin{tabular}[c]{@{}l@{}}QI2. Which one reveals and addresses a learner's knowledge  deficits more clearly?\end{tabular}                                                                                                                   \\ \hline
\end{tabular}%
\Description{This table presents measuring questions and statements used in the expert evaluation of the 'Initial Generation' pipeline and the human evaluation of the end-to-end pipeline of \sysname{} to assess the educational quality of designed dialogue. The table includes three criteria: 'Dynamic,' 'Academic Productivity,' and 'Immersion.' Each criterion is accompanied by statements (with a 7-point Likert Scale) and measuring questions (pairwise comparison) to evaluate the dialogue's characteristics. The table is labeled 'Table 7: Measuring questions and statements for evaluating the educational quality of \sysname{}'s designed dialogue.'}
\caption{Measuring questions used in our expert evaluation of the \textit{Initial Generation} pipeline and statements used in our human evaluation of the end-to-end pipeline of \sysname{} to measure the educational quality of designed dialogue.}
\label{tab:eval_metric}
\end{table*}}{\begin{table*}[]
\resizebox{\textwidth}{!}{%
\begin{tabular}{c|l|l}
\hline
\textbf{Criteria}   & \multicolumn{1}{c|}{\textbf{\begin{tabular}[c]{@{}c@{}}Statement \\ (7-point Likert Scale)\end{tabular}}}                                                                                                                                    & \multicolumn{1}{c}{\textbf{\begin{tabular}[c]{@{}c@{}}Measuring questions \\ (Pairwise Comparison)\end{tabular}}}                                                                                                                              \\ \hline
\textbf{Dynamic} & SD1. The dialogue demonstrates clear and fast turn-taking. & QD1. Which one demonstrates clearer and faster turn-taking?                                                                                                                                                                                    \\
& \begin{tabular}[c]{@{}l@{}}SD2. The dialogue utilizes diverse interaction patterns \\ between a tutor and tutees.\end{tabular}                                                                                                               & \begin{tabular}[c]{@{}l@{}}QD2. Which one utilizes more diverse interaction patterns \\ between a tutor and a tutee?\end{tabular}                                                                                                              \\ \hline
\textbf{\begin{tabular}[c]{@{}c@{}}Academic\\ Productivity\end{tabular}}
& \begin{tabular}[c]{@{}l@{}}SAP1.  The dialogue encourages the learner's \\ cognitive engagement (e.g., asking about what they've \\ learned, asking various types of questions, and inquiring \\ about a student's experiences)\end{tabular} & \begin{tabular}[c]{@{}l@{}}QAP1. Which one encourages the learner's cognitive \\ engagement more? (e.g., asking about what they've \\ learned, asking various types of questions, and inquiring \\ about a student's experiences)\end{tabular} \\
  & \begin{tabular}[c]{@{}l@{}}SAP2. The dialogue prompts a student's metacognitive \\ thinking.\end{tabular}                                                                                                                                    & \begin{tabular}[c]{@{}l@{}}QAP2. Which one prompts a learner's metacognitive \\ thinking more?\end{tabular}                                                                                                                                    \\ \hline
\textbf{Immersion} & \begin{tabular}[c]{@{}l@{}}SI1. The dialogue appears to describe a specific and \\ natural learning situation.\end{tabular}                                                                                                                  & \begin{tabular}[c]{@{}l@{}}QI1. Which one describes a more specific and natural \\ learning situation?\end{tabular}                                                                                                                            \\
  & \begin{tabular}[c]{@{}l@{}}SI2. The dialogue reveals and addresses a learner's \\ knowledge deficits more clearly.\end{tabular}                                                                                                              & \begin{tabular}[c]{@{}l@{}}QI2. Which one reveals and addresses a learner's knowledge \\ deficits more clearly?\end{tabular}                                                                                                                   \\ \hline
\end{tabular}
}
\Description{This table presents measuring questions and statements used in the expert evaluation of the 'Initial Generation' pipeline and the human evaluation of the end-to-end pipeline of \sysname{} to assess the educational quality of designed dialogue. The table includes three criteria: 'Dynamic,' 'Academic Productivity,' and 'Immersion.' Each criterion is accompanied by statements (with a 7-point Likert Scale) and measuring questions (pairwise comparison) to evaluate the dialogue's characteristics. The table is labeled 'Table 7: Measuring questions and statements for evaluating the educational quality of \sysname{}'s designed dialogue.'}
\caption{Measuring questions used in our expert evaluation of the \textit{Initial Generation} pipeline and statements used in our human evaluation of the end-to-end pipeline of \sysname{} to measure the educational quality of designed dialogue. These metrics are developed based on the design recommendations defined in Section 4.2.}
\label{tab:eval_metric}
\end{table*}}

\subsection{Technical Evaluation}
To evaluate whether \sysname{} supports authoring dialogues that meet the design requirements for educational dialogues (Section 4.2), we conducted a technical evaluation focusing on three primary parts: (1) \textit{Initial generation} prompting pipeline, (2) our end-to-end pipeline designed for dialogue authoring, and (3) \textit{Correctness} of the final dialogue. For the human evaluation of two pipeline outputs, we invited four instructors who participated in our user study to evaluate the pedagogical quality of the dialogues using the metrics shown in Table ~\ref{tab:eval_metric}. 


\subsubsection{\textbf{\textit{Initial generation} prompting pipeline evaluation}}
We created a test dataset to explore how dialogues are generated through the Initial Generation prompting pipeline because the pipeline is designed to play the most crucial role in generating quality dialogue. 
Notably, we aim to investigate whether the language and the subject factors affect the quality of the pipeline to test the generalizability of the system for different subjects and languages.   

To do this, we construct our test dataset on two lecture videos. We selected science and mathematics lectures to use: topics for the science lecture (Properties of periodic waves)\footnote{\url{https://www.youtube.com/watch?v=tJW_a6JeXD8\&t=345s}} and topics for mathematics (Linear equation)\footnote{\url{https://www.youtube.com/watch?v=9IUEk9fn2Vs}}, as our target domain is STEM subjects.
In addition, to compare across different languages, we selected Khan Academy videos with transcripts available in both Korean and English. For each subject, we selected one segment of approximately 2-3 minutes for dialogue generation. We generated 32 dialogues that consist of 16 Baseline evaluation dialogues (8 in Korean and 8 in English) and 16 \sysname{} evaluation dialogues (8 in Korean and 8 in English). Detailed test dataset generation process and dialogue examples are in Appendix.

Two evaluators evaluated the Korean dialogues, while the other two who are proficient in reading and listening in English assessed the English dialogues, utilizing the given evaluation metrics (\textit{Measuring questions} column of Table ~\ref{tab:eval_metric}). Each evaluator conducted a pairwise comparison on a set of 32 pairs of dialogues and was asked to choose the dialogue generated by \sysname{} or the Baseline condition. 
Then, we calculated the preference percentage of selecting each condition to provide a comprehensive view of the system comparison. 

\subsubsection{\textbf{\textit{End-to-end} dialogue authoring pipeline evaluation}}
We assessed the final dialogues in two ways. Firstly, we compared the Likert scores to determine which one produced more \textit{Dynamic}, \textit{Academically Productive,} and \textit{Immersive} dialogue. Secondly, we compared the percentage of incorrect responses for each dialogue to evaluate the variations in correctness before and after the instructor's refinement and between the different conditions.

\textit{\textbf{Dynamic, Academically Productivity, and Immersion} evaluation of authored dialogues by \sysname{}.} We collected expert evaluations on 20 dialogues designed during the user study, ten from Baseline and ten from \sysname{}. Each dialogue was evaluated by three or four evaluators, as evaluators did not evaluate the dialogues designed by themselves. The evaluators used the evaluation metrics shown in \emph{Statement} column of Table~\ref{tab:eval_metric}, which consisted of six 7-point Likert-scale questions. 

\textit{\textbf{Correctness evaluation of authored dialogues in both conditions.}}
During the \textit{Refinement} stage, instructors were allowed to make direct modifications. We conducted an evaluation study with four instructors to validate our approach using 48 dialogues from our user study. 24 dialogues were generated before direct modifications by instructors, and 24 were created after modifications. Two instructors each evaluated the same dataset and their respective teaching subjects. We compared the two sets of dialogues to identify how our approach improved \textit{Correctness}.

Based on the definition of typical hallucination~\cite{Zhang2023SirensSI,Ye2023CognitiveMA}, we classified three types of incorrectness that may occur in educational dialogues on a turn-by-turn basis: (1) incorrect case where original numbers or explanations in the transcript were transformed incorrectly, (2) inconsistency observed when the answer deviates from the question from the student (e.g., when a student asks a question about logarithm function but the teacher provides an answer about the exponential function), and (3) inconsistencies observed across multiple turns (e.g., inconsistency in the student's knowledge level).

\begin{figure*}[!h]
\centering 
\includegraphics[width=0.8\textwidth]{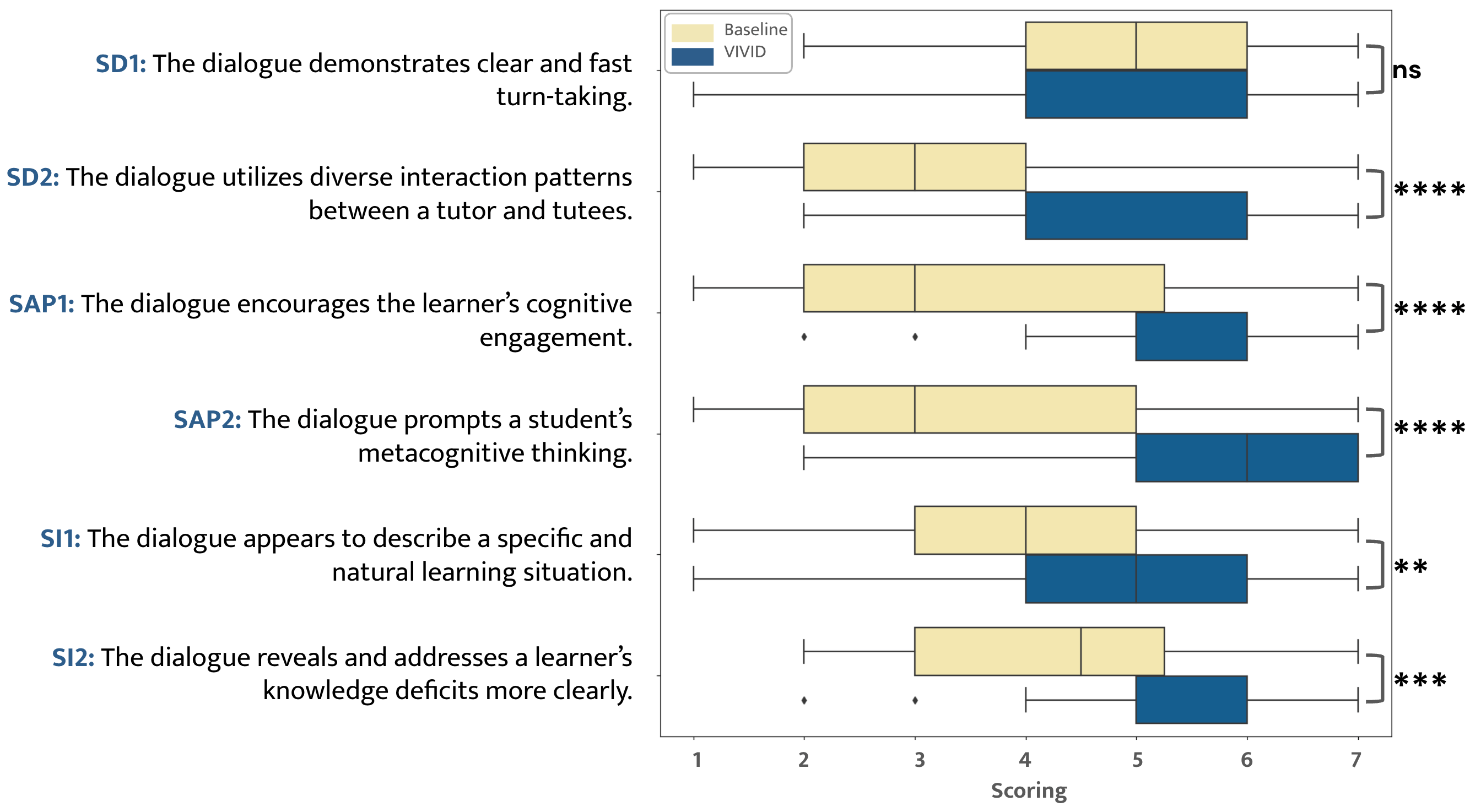} 
\caption{Results of authored dialogues' quality. ****, ***, **, *, and ns indicate significance of \textit{p} <= 0.0001 , 0.0001 < \textit{p} < 0.001, \textit{p} < 0.01, \textit{p} < 0.05, and \textit{p} > 0.05, respectively.}
\label{fig:tech_eval_1}
\end{figure*}

\subsection{Technical Evaluation Results}
The technical evaluation showed that instructors designed significantly higher quality educational dialogues using VIVID compared to Baseline in all criteria except SD1 (Figure~\ref{fig:tech_eval_1}). Our study also found that the \textit{Initial Generation} stage produces better educational dialogues than the Baseline, with the exception of QD1 (Figure~\ref{fig:tech_eval_2}). However, the overall usefulness of each system feature was not significantly high among the instructors as we reported in Section 6.2, so we discussed possible reasons for the gap between the usefulness and quality of dialogues in Section 7.1.


\subsubsection{\textbf{Dynamic, Academically Productivity, and Immersion} evaluation of authored dialogues by \sysname{}}
Technical evaluation results showed that the instructors created significantly better educational dialogues with \sysname{} than the baseline. As shown in Figure~\ref{fig:tech_eval_1}, the dialogues designed through the entire pipeline of \sysname{} were rated significantly higher in quality in all aspects, except for SD1, compared to the baseline. The most significant findings were shown in SD2 (\sysname{}: M= 5.11, SD= 1.45, Baseline: M= 3.4, SD= 1.73), SAP1 (\sysname{}: M= 5.47, SD= 1.13, Baseline: M= 3.7, SD= 1.9), SAP2 (\sysname{}: M= 5.64, SD= 1.4, Baseline: M= 3.3, SD= 2, p <= 1.00e-04, Wilcoxon signed-rank test). In other words, dialogues authored by the end-to-end pipeline of \sysname{} better described the metacognitive and cognitive activities of direct learners and consisted of more diverse patterns than the baseline. Difference of SI2 (p <= 0.001) and SI1 (p <= 0.01) also showed significance. The result implies that the dialogues authored with \sysname{} described a more natural learning situation and the direct learner’s knowledge deficit better than the baseline.


\subsubsection{\textbf{Correctness} evaluation of authored dialogues in both conditions.}
\begin{figure}[h]
\centering 
\includegraphics[width=0.4\textwidth]{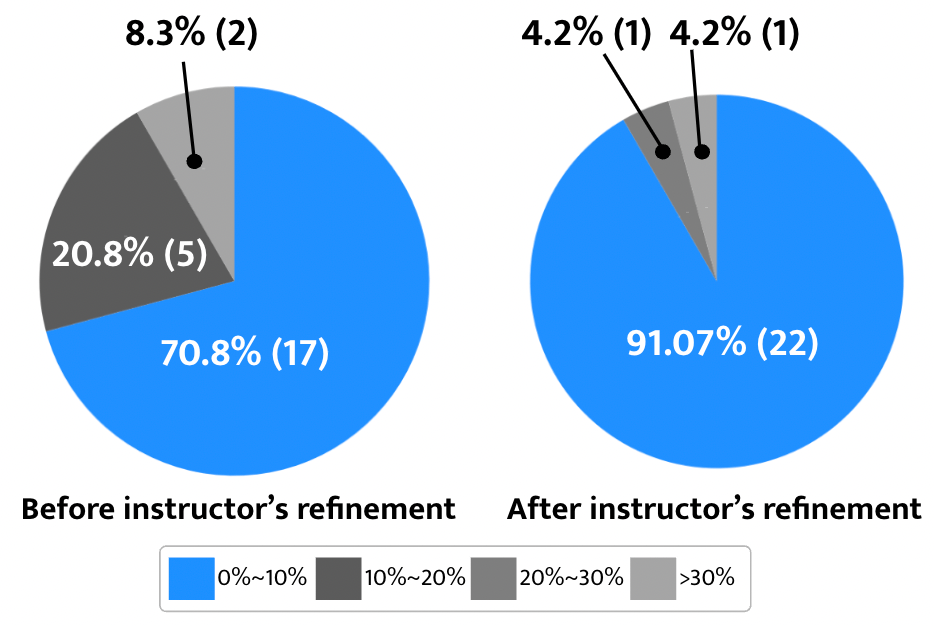} 
\caption{Human evaluation results on \textit{Correctness}. The figure illustrates how the instructor's refinement has affected the correctness.}
\Description{Correctness}
\label{fig:correctness}
\end{figure}

\begin{figure}[h] 
\centering 
\includegraphics[width=0.5\textwidth]{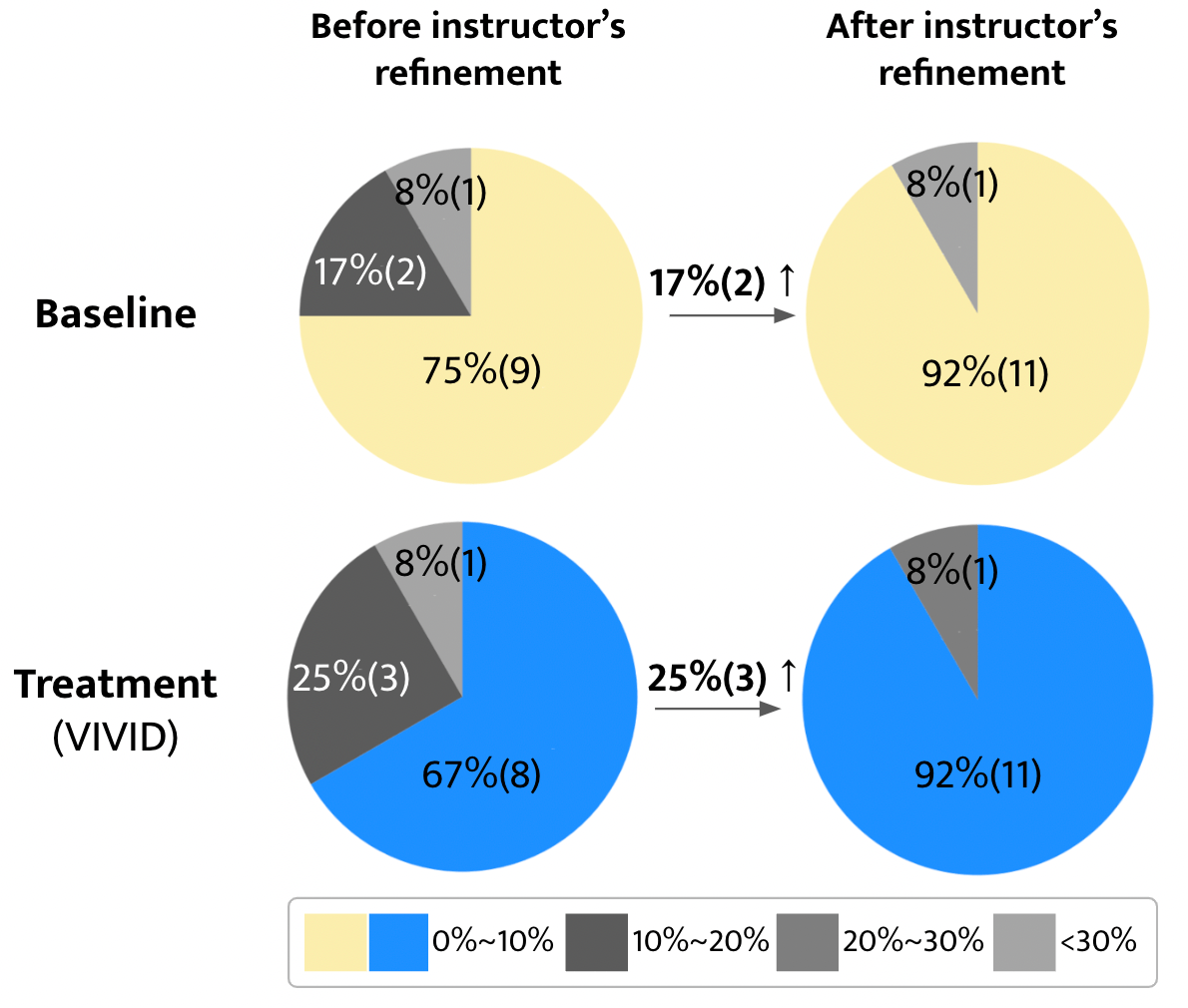} 
\caption{Human evaluation results on \textit{Correctness} according to condition. This figure shows bigger changes in \sysname{} and high correctness of final dialogues in both conditions. The questionnaire used in this evaluation is from Table~\ref{tab:eval_metric}.}
\Description{Correctness}
\label{fig:correctness_2}
\end{figure}

We analyzed the percentage of turns with errors in each dialogue. As shown in Figure~\ref{fig:correctness}, after the modification, the total incorrectness rate of 0\%-10\% increased from 71\% (17) to 92\% (22). 
Before modification, VIVID generated more incorrect dialogues than Baseline because VIVID had to consider more details about the direct learner's understanding states when generating dialogue. 
After modification, the percentage of VIVID’s incorrect dialogues in the 0-10\% range increased from 67\% (8) to 92\% (11), while Baseline increased from 75\% (9) to 92\% (11) (Figure~\ref{fig:correctness_2}).
These results indicate that \sysname{} improved correctness better than Baseline, and the instructor's refinement produced high-correctness final dialogues in both conditions. Additionally, we calculated the percentage of exceptional incorrect dialogues due to transcript errors (e.g., absence of essential conditions like `x < 0', incorrect concept definition). Before the instructor made corrections, 25\% of the entire dialogue resulted from incorrect transcripts. Even after the instructor's refinement, around 17\% persisted, especially due to the absence of essential conditions in math dialogues.

\begin{figure*}[!h] 
\centering 
\includegraphics[width=\textwidth]{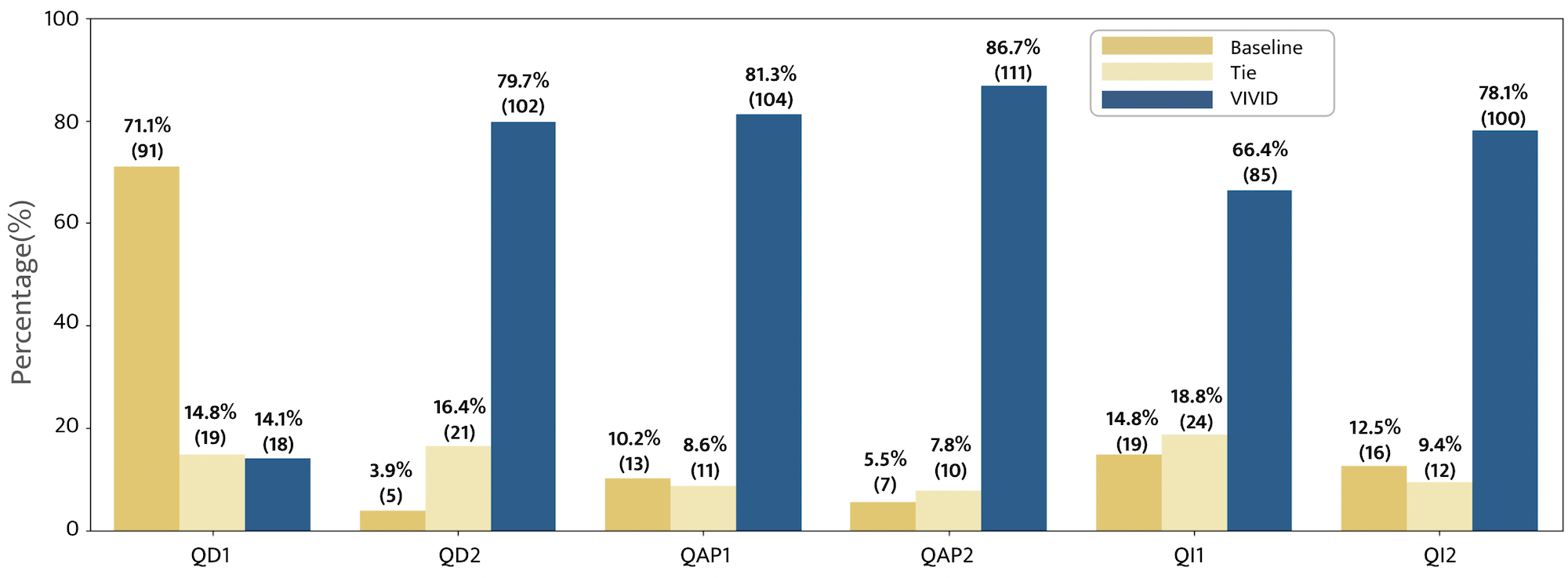} 
\caption{Human evaluation results on the six questions listed in Table~\ref{tab:eval_metric}. Four instructors evaluated dialogue sets, and each instructor conducted a pairwise comparison on a set of 32 pairs of dialogues. Except for QD1, \sysname{} outperformed the Baseline in the other five metrics.}
\Description{Human evaluation results on the six questions. The questions used are listed in Table~\ref{tab:eval_metric}}
\label{fig:tech_eval_2}
\end{figure*}

\subsubsection{\textit{Initial Generation} pipeline evaluation.}
The dialogues generated by \sysname{}'s \textit{Initial Generation} pipeline were rated higher in quality compared to the corresponding baseline in terms of all metrics listed in Table~\ref{tab:eval_metric}, except QD1. 
The most significant difference (Baseline: 5.5\%, \sysname{}: 86.7\%) was on QAP2 (Table~\ref{tab:eval_metric}) as in the end-to-end pipeline (Section 6.4.1), indicating that \sysname{} generates initial dialogues that effectively reflect a direct learner's metacognitive activity (Figure~\ref{fig:tech_eval_2}).
Figure~\ref{fig:tech_eval_2} shows that QD2, QAP1, QI1, and QI2 had over a 50\% difference between \sysname{} and Baseline except for QD1 (Baseline: 71.01\%, \sysname{}: 14.1\%). We discussed the issue of poor quality for QD1 in Section 7.1.

\begin{figure*}[!h] 
\centering 
\includegraphics[width=0.8\textwidth]{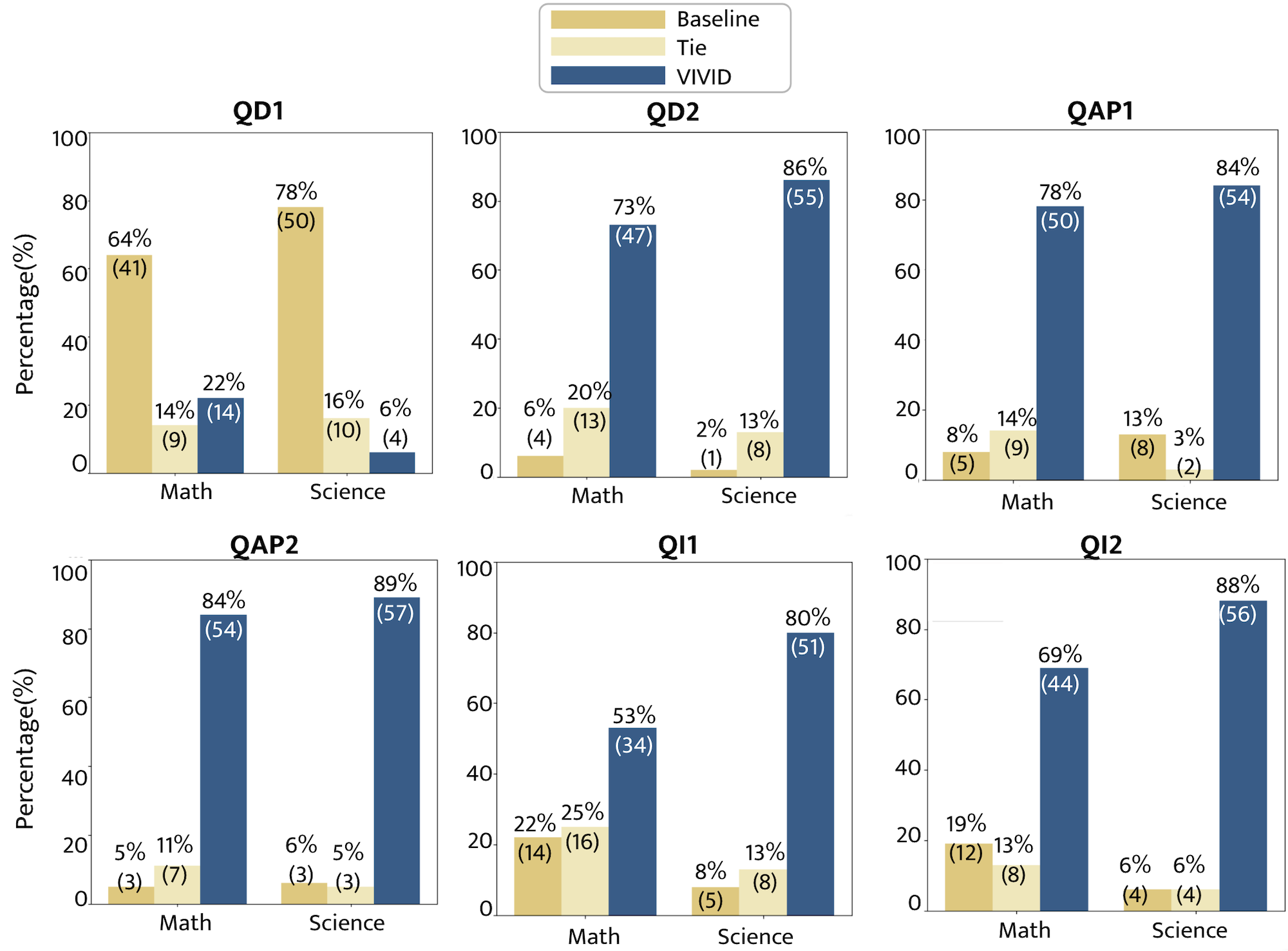} 
\caption{Results of dialogue quality by subject (Science, Math). The questions used are listed in Table~\ref{tab:eval_metric}. Except for QD1, VIVID outperformed the Baseline
in the other five metrics.}
\Description{Results of dialogue quality by subject (Science, Math)}
\label{fig:tech_eval_2_subject}
\end{figure*}

\textbf{Dialogue quality difference by subject (Science, Math)}
As shown in Figure~\ref{fig:tech_eval_2_subject}, the most significant difference between Baseline (2\%) and \sysname{} (86\%) was observed in the QD2 criterion, which is about diverse interaction patterns, in science dialogues. This suggests that \sysname{} effectively utilizes diverse dialogue patterns between the tutor and the direct learner, regardless of the language used. 

Regarding math videos, the evaluation metric with the largest difference between Baseline (5\%) and \sysname{} (84\%) was QAP2 (Table~\ref{tab:eval_metric}). This indicates that \sysname{} is particularly effective in designing a dialogue that encourages metacognitive speech from a direct learner, regardless of the language used. On the other hand, QAP1 (Baseline: 13\%, \sysname{}: 84\%) and QI1 (Baseline: 8\%, \sysname{}: 80\%) showed smaller differences, but were still significant. In both subjects, evaluators preferred the dialogues generated by the Baseline in terms of QD1 (Table~\ref{tab:eval_metric}) which is about verbosity. 

\begin{figure*}[!h] 
\centering 
\includegraphics[width=0.8\textwidth]{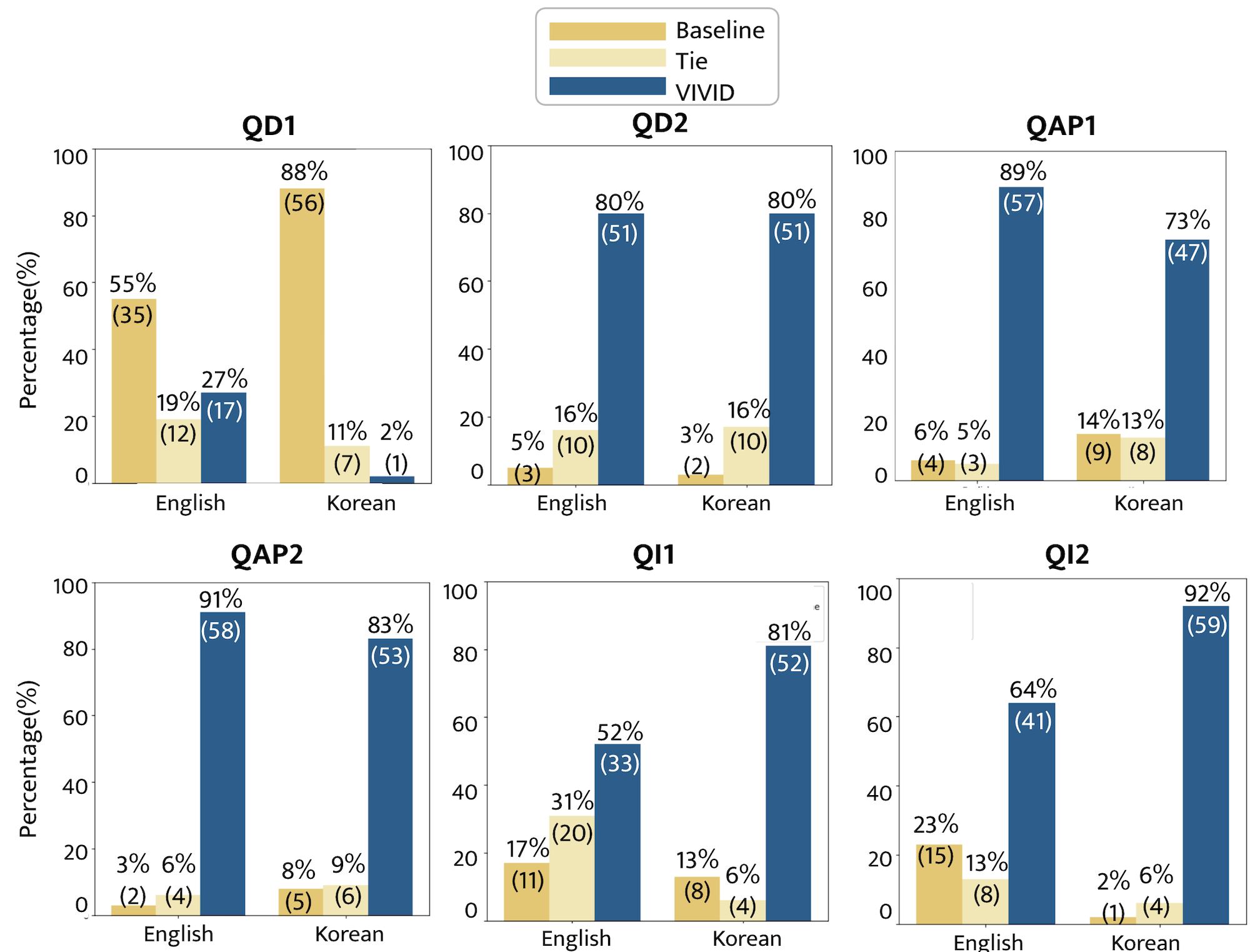} 
\caption{Results of generated dialogue quality by language (Korean, English). The questions used are listed in Table~\ref{tab:eval_metric}. Except for QD1, VIVID outperformed the Baseline
in the other five metrics.}
\Description{Results of generated dialogue quality by language (Korean, English).}
\label{fig:tech_eval_2_lang}
\end{figure*}

\textbf{Dialogue quality difference by language (English, Korean)}
As shown in Figure~\ref{fig:tech_eval_2_lang}, QAP2 (Table~\ref{tab:eval_metric}) had the most significant difference between Baseline (3\%) and \sysname{} (91\%) in English. This suggests that despite the subject, \sysname{} effectively created dialogues that elicited metacognitive activities from the instructor to the learner when the dialogues were converted from English to English.
Yet, the Baseline performed better in terms of QD1 better than \sysname{}, both in English (Baseline: 41\%, \sysname{}: 14\%) and Korean (Baseline: 50\%, \sysname{}: 4\%) as in the results of dialogue quality difference by subject.

When converting Korean lecture into Korean dialogue, \sysname{} showed a significant contrast between Baseline (6\%) and \sysname{} (92\%) in the QI2 criterion while QAP1 had the least difference between Baseline (31\%) and \sysname{} (51\%). 
This indicates that \sysname{} effectively represented the learner's knowledge gaps directly and clearly, and depicted the process of addressing these difficulties in the dialogue, regardless of the subject. 

\section{Discussion}

In this section, we discussed how to improve explainability, controllability, and verbosity for better utility, \sysname{}'s potential beyond lecture videos, customizability for learners, and its generalizability.


\subsection{Human-AI Interaction Design for \sysname{}}
The technical evaluation showed that dialogues created using \sysname{} were significantly better than Baseline in five out of six criteria (Figure~\ref{fig:tech_eval_1}), excluding fast turn-taking. Similarly, the quality of the \sysname{}-generated dialogues during \textit{Initial Generation} phase was rated significantly higher. However, there was no significant difference in the instructors' perceived efficiency and usefulness of VIVID compared to Baseline (Figure~\ref{fig:post_survey}).

Despite the positive results for \sysname{}, the overall usefulness of each system feature was relatively low among the instructors (Figure~\ref{fig:post_survey}). We attribute this to two reasons: (1) low explainability and informativeness of dialogue design described in the \textit{highlighting} feature and \textit{dialogue cards} and (2) low controllability of the \textit{laboratory} feature. 
Thus, we suggest three improvements:

\begin{itemize}
    \item \textbf{Enhancing Explainability}: 
    The \textit{highlighting} feature and \textit{dialogue cards} in \sysname{} need to offer greater explainability to the instructors. P7 highlighted that having prior knowledge of each feature's exact functionality could have led to more frequent and appropriate usage, potentially resulting in a higher level of satisfaction with the system. Notably, there is a need to investigate the types of information instructors require to effectively discern the diversity among learners and pedagogical dialogue patterns. We observed substantial differences between instructors in their ability to recognize differences in direct learners’ understanding levels and how these differences are reflected in dialogue structure.
    
    \item \textbf{Providing Fine-grained Controllability}: 
    Enhancing controllability and providing granular modifications on the \textit{laboratory} feature can improve instructors' workflow. In our user study, we observed that instructors exhibited varying expectations for the modified versions offered by the \textit{laboratory}, and they tended to rate usability lower when their expectations were not met. The improved version of \textit{laboratory} feature could support the instructors in determining and expressing what features they would expect in the revised versions of the dialogue. For instance, enabling instructors to select elements, such as diverse versions of examples, questions, or versions with added prior knowledge with interactive guidance, could increase the perceived usefulness of the feature.

    \item \textbf{Improving Verbosity}:
    One unexpected downside was that the generated dialogues were perceived to be verbose, which is likely due to LLM's tendency to produce long text. This issue could be addressed by revising the prompting pipeline to limit the length of utterances and dialogues generated, which we leave as future work.
\end{itemize}

\subsection{Potential Applications beyond the Video Lecture Context} 
In the educational context, dialogues can serve multiple roles, extending beyond the mere transmission of factual knowledge. In our user study, several instructors highlighted the adaptability of our dialogue design pipeline, suggesting its potential application in diverse learning contexts, instructional materials, and various learning stages. For instance, P7 proposed utilizing our dialogue design pipeline to use dialogues for the learner's review, or to use dialogue as a means to diagnose the learner's misconceptions by providing a dialogue in which the direct learner presents misconceptions. 

Furthermore, \sysname{} and its process of transforming lectures into a dyadic format may take the role of a valuable active learning tool. Our dialogue design pipeline can be utilized in formulating questions in a dialogue format for learners and provide interactive guidance for the students' self-learning process using digital textbooks or in flipped learning settings. Learners can gain a better understanding of complex concepts by analyzing educational content and exploring effective teaching strategies.


\subsection{Customizable \sysname{} for learners}
\sysname{} is a system that supports instructors in transforming their lecture videos into educational dialogues in text format. Yet, it is important to consider how these dialogues can be seamlessly incorporated into the video learning environment (VLE) to enrich learners' experiences and optimize learning outcomes.
We can integrate text-format dialogue into the VLE by delivering it in voice and text modes together, utilizing the VLE's multi-modality. For instance, the dialogue can be converted into human-like speech and played alongside the corresponding lecture clip by replacing the original explanations with dialogue. Furthermore, vicarious learners can simultaneously explore multimodal dialogue incorporating formulas in the lecture within a chat-like interface.

While \sysname{}, designed for instructors, utilizes data pertinent to the levels of a vicarious learner group considered by instructors, it is limited when incorporating teaching strategies, like transfer learning (DR2 in Section 4.2) and personalization of dialogue, which demand each vicarious learner’s data, such as prior knowledge, personal background, and current understanding state.
We believe that \sysname{} can be extended to collect data from vicarious learners by using a multi-modal representation of vicarious dialogue. This would enable customized modeling of direct learners, effective transfer learning, and personalization to vicarious learning. For instance, we can collect the data for generating personalized dialogue by requiring learners to click on challenging elements such as formulas or explanations within a lecture as they watch it. Therefore, future work should expand \sysname{} to include learners and evaluate dialogues from the learner-centered criteria, such as engagement and learning gain.

\subsection{Generalizability of \sysname{}}
Even when different lectures cover the same concept, variables such as material modalities, style of delivery, and language affect how a learner perceives and understands new knowledge. We found that instructors tend to adjust dialogue to fit their teaching style when the teaching style in the lecture differs from their preference. To enable instructors to use lecture videos of any teaching style and match them with their intended outcomes, it is necessary to explore a solution for converting dialogue, which includes a preprocessing step for scripting before the \textit{Initial Generation} phase. To design dialogues based on lectures with varying teaching styles, \sysname{} needs to preprocess the lecture material to isolate core concepts, understand the instructor's intention, and transform the knowledge into a personalized format that matches the user's preferred teaching style.

Moreover, it is important to determine which lecture segments and lengths are suitable for a dialogue style. As P3 said, certain contents or subjects may be more suitable for dialogue formats to help learners better understand relatively complex concepts or examples. Further, we observed in our technical evaluation that the dialogue generation had varying degrees of improvement depending on the subject matter. Thus, enhancing the advantages of dialogue format can be achieved by understanding and reflecting on the differing effects of dialogue format between subjects in dialogue design.



\subsection{Limitations and Future Work}
We acknowledge several limitations in our current study. 
Firstly, the knowledge progression of the direct learner in the dialogue was not one-sided in \sysname{}. 
\sysname{} didn't consider prerequisite relationships to create diverse dialogues (Section 5.1.2). Yet, some dialogues depicted direct learners initially understanding a concept but later appearing to lack understanding. Thus, redesigning the knowledge state setting pipeline is needed to maintain consistent knowledge levels and prevent reverse progression.
Secondly, our experiments involved instructors designing dialogues for only a single segment within a lecture. However, the generated dialogues are influenced by factors such as the length of the selected segment, the type of content, and the subject. To explore \sysname{}'s use cases more deeply, it is necessary to conduct experiments under a more diverse set of conditions.

\section{Conclusion}
We present design recommendations from an extensive literature review and insights gathered during a design workshop. These recommendations are aimed at facilitating the creation of high-quality educational dialogues. To put these guidelines into practice, we have developed \sysname{}, a web application designed to assist instructors in authoring pedagogical dialogues from their monologue-style lecture videos. Through our technical evaluation and user study, we found that instructors can consider important factors in dialogue design effectively, generating \textit{Dynamic, Pedagogically productive, Immersive}, and \textit{Correct} dialogues. We hope VIVID helps create more engaging lecture videos, providing a personalized learning experience for online students.

\begin{acks}
This work was supported by Elice. This work was also supported by Institute of Information \& Communications Technology Planning \& Evaluation (IITP) grant funded by the Korea government (MSIT) (No.2021-0-01347, Video Interaction Technologies Using Object-Oriented Video Modeling). We thank all of the members of KIXLAB (KAIST Interaction Lab) for their discussions and constructive feedback. 
\end{acks}

\balance
\bibliographystyle{ACM-Reference-Format}
\bibliography{main}

\clearpage
\appendix

\section{Workshop Details}
\subsection{Subjects and Lesson Content Used In Workshop}
\aptLtoX{\begin{table*}[h!]
\begin{tabular}{l|c|l}
\hline
\multicolumn{1}{c|}{\textbf{Group}} & \textbf{Subject} & \multicolumn{1}{c}{\textbf{Main lecture content}}                                                                                                                                                                                                                      \\ \hline
1 (P1, P2)                          & Math             & Composite functions and inverse functions                                                                                                                                                                                                                              \\ \hline
2 (P3, P4)                          & Science          & Phases of the moon and the reasons behind these lunar phases                                                                                                                                                                                                           \\ \hline
3 (P5, P6)                          & Math             & Concept of unit vectors and their alignment with a given vector                                                                                                                                                                                                        \\ \hline
4 (P7, P8)                          & Science          & \begin{tabular}[c]{@{}l@{}}Einstein's General Theory of Relativity, covering concepts such as the warping of spacetime but massive  celestial bodies, gravitational lensing, time dilation due to gravity, and phenomena associated with black  holes\end{tabular} \\ \hline
5 (P9, P10)                         & Math             & Concepts of radical (square root) functions and rational (fractional) functions                                                                                                                                                                                        \\ \hline
6 (P11, P12)                        & Math             & \begin{tabular}[c]{@{}l@{}}Process of transforming a quadratic equation into a perfect square trinomial and then using square roots  to find the solutions\end{tabular}                                                                                              \\ \hline
7 (P13, P14)                        & Math             & \begin{tabular}[c]{@{}l@{}}Method of expressing a third line passing through the intersection of two given lines and determining the  equation of a line, even with an unknown slope, passing through a specified point\end{tabular}                                 \\ \hline
8 (P15)                             & Math             & Classification of integers based on the remainders when divided by a positive integer                                                                                                                                                                                  \\ \hline
\end{tabular}%
\caption{Subjects and lesson contents of the lecture that were addressed in the workshop by each group.}
\label{tab:workshop_lecture_info}
\end{table*}}{\begin{table*}[h!]
\resizebox{\textwidth}{!}{%
\begin{tabular}{l|c|l}
\hline
\multicolumn{1}{c|}{\textbf{Group}} & \textbf{Subject} & \multicolumn{1}{c}{\textbf{Main lecture content}}                                                                                                                                                                                                                      \\ \hline
1 (P1, P2)                          & Math             & Composite functions and inverse functions                                                                                                                                                                                                                              \\ \hline
2 (P3, P4)                          & Science          & Phases of the moon and the reasons behind these lunar phases                                                                                                                                                                                                           \\ \hline
3 (P5, P6)                          & Math             & Concept of unit vectors and their alignment with a given vector                                                                                                                                                                                                        \\ \hline
4 (P7, P8)                          & Science          & \begin{tabular}[c]{@{}l@{}}Einstein's General Theory of Relativity, covering concepts such as the warping of spacetime but massive \\ celestial bodies, gravitational lensing, time dilation due to gravity, and phenomena associated with black \\ holes\end{tabular} \\ \hline
5 (P9, P10)                         & Math             & Concepts of radical (square root) functions and rational (fractional) functions                                                                                                                                                                                        \\ \hline
6 (P11, P12)                        & Math             & \begin{tabular}[c]{@{}l@{}}Process of transforming a quadratic equation into a perfect square trinomial and then using square roots \\ to find the solutions\end{tabular}                                                                                              \\ \hline
7 (P13, P14)                        & Math             & \begin{tabular}[c]{@{}l@{}}Method of expressing a third line passing through the intersection of two given lines and determining the \\ equation of a line, even with an unknown slope, passing through a specified point\end{tabular}                                 \\ \hline
8 (P15)                             & Math             & Classification of integers based on the remainders when divided by a positive integer                                                                                                                                                                                  \\ \hline
\end{tabular}%
}
\caption{Subjects and lesson content of the lecture that were addressed in the workshop by each group.}
\label{tab:workshop_lecture_info}
\end{table*}}
Table~\ref{tab:workshop_lecture_info} indicates the subjects and lesson contents used in our workshop.

\subsection{Teaching Strategies for Designing Pedagogically Effective Dialogue.}
\aptLtoX{\begin{table*}[h]
\begin{tabular}{c|c|c|ll}
\hline
\textbf{By initiative} & \begin{tabular}[c]{@{}c@{}}Key strategies that  can be effective to vicarious learners\end{tabular} & Description                                                                                                                                                                                                                                                                                                                              & \multicolumn{2}{c}{\begin{tabular}[c]{@{}c@{}}Example Dialogues  between a tutor and a tutee\end{tabular}}                                                                                                                                                                                                                                                                                                                                                                                                                                        \\ \hline
\multirow{3}{*}{Tutor} & \textbf{Cognitive conflict}    & \begin{tabular}[c]{@{}c@{}}It is a teaching strategy that examines  the learner's prior knowledge, creates a  mismatch situation that causes conflict,  and then helps the learners to see that  his or her understanding is incorrect.\end{tabular}                                                                             & \multicolumn{1}{l|}{\textbf{\begin{tabular}[c]{@{}l@{}}Tutor\\ \\ \\ \\ Tutee\\ Tutor\end{tabular}}}          & \begin{tabular}[c]{@{}l@{}}Absolutely, that's the usual method. But let  me throw a curveball. What if I told you that  solving them using a different approach might  lead to a different solution? Really? I thought there was only one way to solve  equations. That's what we're here to explore! Let's try this. Instead of isolating x right away, …\end{tabular}                                              \\ \cline{2-5} 
                       & \textbf{Metacognitive  prompting}                                                                     & \begin{tabular}[c]{@{}c@{}}It orients learners towards higher-level  strategies (e.g., goal-setting, planning, monitoring, evaluation, reflection).  It includes an instructor’s utterances that  encourage the learner to express their current  level of understanding or articulate their thought process.\end{tabular} & \multicolumn{1}{l|}{\textbf{\begin{tabular}[c]{@{}l@{}}Tutor\\ \\ \\ \\ \\ Tutee\\ \\ \\ Tutor\end{tabular}}} & \begin{tabular}[c]{@{}l@{}}Got it. How about we take a slightly different approach this time? Before you jump into solving,  let’s start by identifying what the problem is asking.  Can you read the question and tell me what this  question is requesting? Sure. It’s asking me to solve for the sum of ‘x’  and ‘y’ in the equation. That’s a nice interpretation, but let’s take a closer look.\end{tabular} \\ \cline{2-5} 
                       & \textbf{Cognitive prompting}                                                                          & \begin{tabular}[c]{@{}c@{}}It engages learners in lower-level strategies  (e.g., organization, rehearsal, elaboration).  It includes the instructor's utterances that  prompt the learner to talk about what they are  learning or to draw out the learner's  prior knowledge and personal experiences.\end{tabular}           & \multicolumn{1}{l|}{\textbf{\begin{tabular}[c]{@{}l@{}}Tutor\\ \\ \\ Tutee\end{tabular}}}                     & \begin{tabular}[c]{@{}l@{}}As you work through an equation, think about the  basic operations you've learned. Can you explain how  these operations are helping you manipulate this equation? Sure. When there’s addition on one side, I subtract to  balance it out. And if it’s multiplication, I divide to get ‘x’ by itself.\end{tabular}                                                                            \\ \hline
Tutee                  & \textbf{Spontaneous deep-level reasoning question}                                                    & \begin{tabular}[c]{@{}c@{}}It refers to starting a conversation where  the learner spontaneously asks deep-level  reasoning questions that help them better  understand and engage in critical thinking.\end{tabular}                                                                                                              & \multicolumn{1}{l|}{\begin{tabular}[c]{@{}l@{}}\textbf{Tutee}\\ \\ \textbf{Tutor}\end{tabular}}                        & \begin{tabular}[c]{@{}l@{}}\textit{How can a manufacturer increase the speed of the computer?} \textit{What can they do to make it faster?} \textit{Well, one thing manufacturers do is increase the 
clock speed}  \textit{of the computer.}\end{tabular}                                                                                                                                                                                                \\ \hline
\end{tabular}%
\Description{This table presents four teaching strategies for pedagogically effective dialogue: Cognitive conflict, Metacognitive prompting, Cognitive prompting, and Spontaneous deep-level reasoning question. The table consists of five columns: 'By initiative,' 'Key strategies that can be effective to vicarious learners,' 'Description,' 'Example Dialogues between a tutor and a tutee (Tutor),' and 'Example Dialogues between a tutor and a tutee (Tutee).' Under 'By initiative,' the table distinguishes between 'Tutor' and 'Tutee' as the initiators of these strategies. The 'Key strategies that can be effective to vicarious learners' column outlines the teaching strategies. The 'Description' column provides an explanation of each strategy. The last two columns showcase example dialogues between a tutor and a tutee to illustrate each strategy. The table is titled 'Table 4: Four teaching strategies for pedagogically effective dialogue.}
\caption{Four teaching strategies for pedagogically effective dialogue: Cognitive conflict, Metacognitive prompting, Cognitive prompting, and Spontaneous deep-level reasoning question.}
\label{tab:teaching-strategies}
\end{table*}}{\begin{table*}[h]
\resizebox{\textwidth}{!}{%
\begin{tabular}{c|c|c|ll}
\hline
\textbf{By initiative} & \begin{tabular}[c]{@{}c@{}}Key strategies that \\ can be effective to vicarious learners\end{tabular} & Description                                                                                                                                                                                                                                                                                                                              & \multicolumn{2}{c}{\begin{tabular}[c]{@{}c@{}}Example Dialogues \\ between a tutor and a tutee\end{tabular}}                                                                                                                                                                                                                                                                                                                                                                                                                                        \\ \hline
\multirow{3}{*}{Tutor} & \textbf{Cognitive conflict}                                                                           & \begin{tabular}[c]{@{}c@{}}It is a teaching strategy that examines \\ the learner's prior knowledge, creates a \\ mismatch situation that causes conflict, \\ and then helps the learners to see that \\ his or her understanding is incorrect.\end{tabular}                                                                             & \multicolumn{1}{l|}{\textbf{\begin{tabular}[c]{@{}l@{}}Tutor\\ \\ \\ \\ Tutee\\ Tutor\end{tabular}}}          & \begin{tabular}[c]{@{}l@{}}Absolutely, that's the usual method. But let \\ me throw a curveball. What if I told you that \\ solving them using a different approach might \\ lead to a different solution?\\ Really? I thought there was only one way to solve \\ equations.\\ That's what we're here to explore! Let's try this. \\ Instead of isolating x right away, …\end{tabular}                                              \\ \cline{2-5} 
                       & \textbf{Metacognitive  prompting}                                                                     & \begin{tabular}[c]{@{}c@{}}It orients learners towards higher-level \\ strategies (e.g., goal-setting, planning, \\ monitoring, evaluation, reflection). \\ It includes an instructor’s utterances that \\ encourage the learner to express their current \\ level of understanding or articulate their \\ thought process.\end{tabular} & \multicolumn{1}{l|}{\textbf{\begin{tabular}[c]{@{}l@{}}Tutor\\ \\ \\ \\ \\ Tutee\\ \\ \\ Tutor\end{tabular}}} & \begin{tabular}[c]{@{}l@{}}Got it. How about we take a slightly different \\ approach this time? Before you jump into solving, \\ let’s start by identifying what the problem is asking. \\ Can you read the question and tell me what this \\ question is requesting?\\ Sure. It’s asking me to solve for the sum of ‘x’ \\ and ‘y’ in the equation.\\ \\ That’s a nice interpretation, but let’s take a closer look.\end{tabular} \\ \cline{2-5} 
                       & \textbf{Cognitive prompting}                                                                          & \begin{tabular}[c]{@{}c@{}}It engages learners in lower-level strategies \\ (e.g., organization, rehearsal, elaboration). \\ It includes the instructor's utterances that \\ prompt the learner to talk about what they are \\ learning or to draw out the learner's \\ prior knowledge and personal experiences.\end{tabular}           & \multicolumn{1}{l|}{\textbf{\begin{tabular}[c]{@{}l@{}}Tutor\\ \\ \\ Tutee\end{tabular}}}                     & \begin{tabular}[c]{@{}l@{}}As you work through an equation, think about the \\ basic operations you've learned. Can you explain how \\ these operations are helping you manipulate this equation?\\ Sure. When there’s addition on one side, I subtract to \\ balance it out. And if it’s multiplication, I divide to get ‘\\ x’ by itself.\end{tabular}                                                                            \\ \hline
Tutee                  & \textbf{Spontaneous deep-level reasoning question}                                                    & \begin{tabular}[c]{@{}c@{}}It refers to starting a conversation where \\ the learner spontaneously asks deep-level \\ reasoning questions that help them better \\ understand and engage in critical thinking.\end{tabular}                                                                                                              & \multicolumn{1}{l|}{\begin{tabular}[c]{@{}l@{}}\textbf{Tutee}\\ \\ \textbf{Tutor}\end{tabular}}                        & \begin{tabular}[c]{@{}l@{}}\textit{How can a manufacturer increase the speed of the computer?}\\ \textit{What can they do to make it faster?}\\ \textit{Well, one thing manufacturers do is increase the 
clock speed} \\ \textit{of the computer.}\end{tabular}                                                                                                                                                                                                \\ \hline
\end{tabular}%
}
\Description{This table presents four teaching strategies for pedagogically effective dialogue: Cognitive conflict, Metacognitive prompting, Cognitive prompting, and Spontaneous deep-level reasoning question. The table consists of five columns: 'By initiative,' 'Key strategies that can be effective to vicarious learners,' 'Description,' 'Example Dialogues between a tutor and a tutee (Tutor),' and 'Example Dialogues between a tutor and a tutee (Tutee).' Under 'By initiative,' the table distinguishes between 'Tutor' and 'Tutee' as the initiators of these strategies. The 'Key strategies that can be effective to vicarious learners' column outlines the teaching strategies. The 'Description' column provides an explanation of each strategy. The last two columns showcase example dialogues between a tutor and a tutee to illustrate each strategy. The table is titled 'Table 4: Four teaching strategies for pedagogically effective dialogue.}
\caption{Four teaching strategies for pedagogically effective vicarious dialogues: Cognitive conflict, Metacognitive prompting, Cognitive prompting, and Spontaneous deep-level reasoning question.}
\label{tab:teaching-strategies}
\end{table*}}
Table~\ref{tab:teaching-strategies} indicates the teaching strategies for designing pedagogically effective dialogue. 

\section{Dialogue Generation Examples}

\subsection{Evaluation Dataset Generation Process}
\begin{figure*}[!h] 
\centering 
\includegraphics[width=0.4\textwidth]{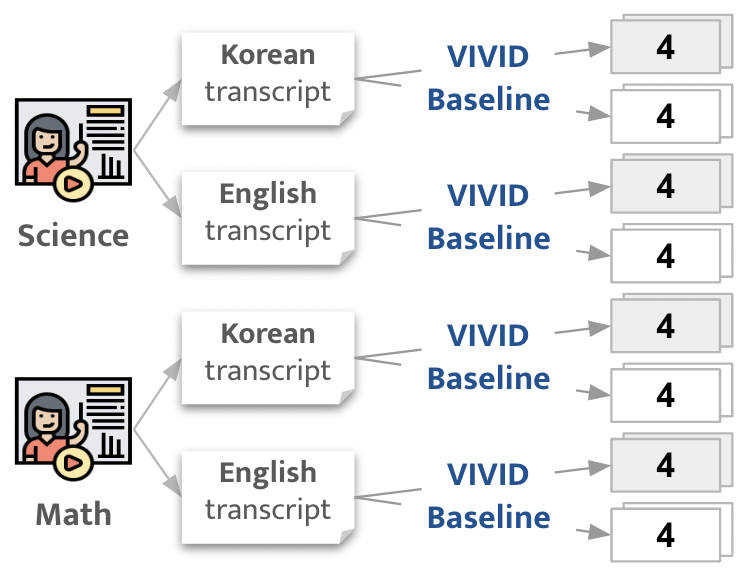} 
\caption{Evaluation Dataset Generation Process.}
\Description{Evaluation Dataset Generation Process.}
\label{fig:test_data}
\end{figure*}

Figure~\ref{fig:test_data} indicates the evaluation dataset generation process. 

\subsection{Transcript Example}
\begin{figure*}[!h]
\centering 
\includegraphics[width=\textwidth]{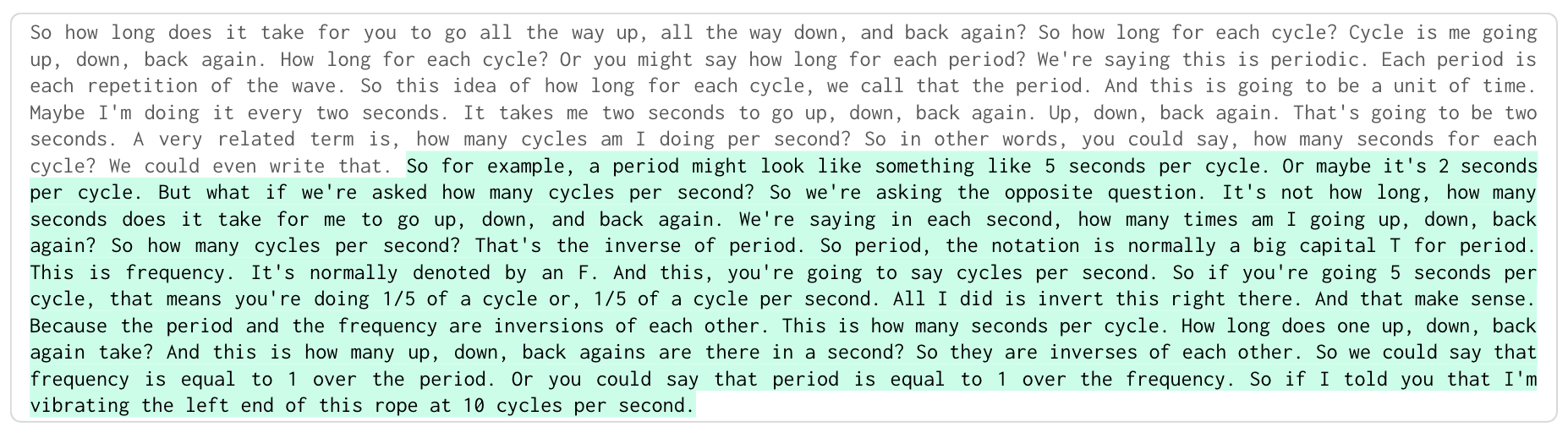} 
\caption{A transcript that is used to generate dialogue data for technical evaluation study (Figure~\ref{fig:dialog_ex1}, ~\ref{fig:dialog_ex2}, ~\ref{fig:dialog_ex3},  ~\ref{fig:dialog_ex4}). This transcript is from a physics lecture used in the technical evaluation study. The physics lecture from which this script was derived supports Korean and English scripts, enabling the generation of dialogues in both languages, and this is an English transcript. The authors highlighted the \textit{green part} as potentially difficult for vicarious learners to understand when generating dialogue data.}
\label{fig:transcript}
\end{figure*}

The green section indicates the area highlighted by the authors as potentially challenging for vicarious learners to understand. \sysname{} creates a direct learner who lacks knowledge of this green area.


\subsection{Example 1}

\begin{figure*}[!h]
\centering 
\includegraphics[width=\textwidth]{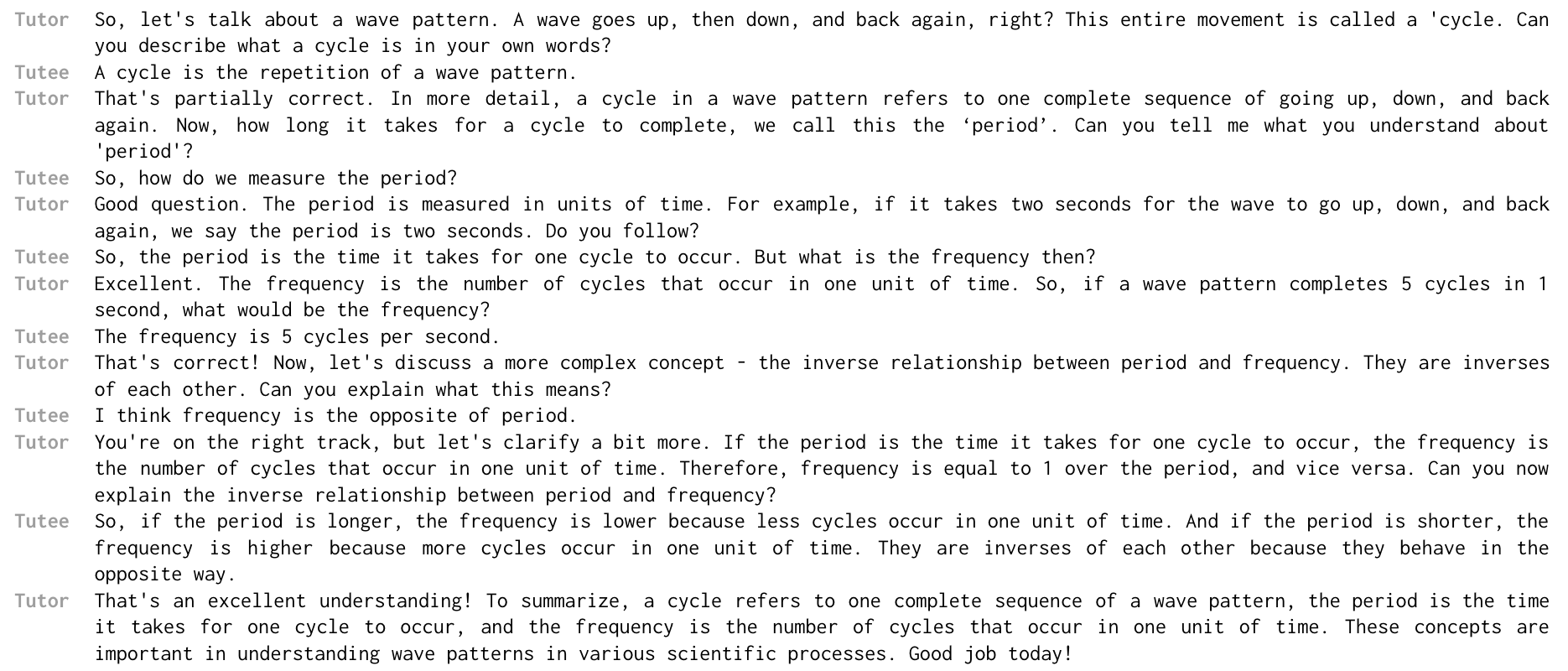} 
\caption{An example of technical evaluation dataset. This dialogue is one of the 16 examples generated by \textbf{\sysname{}} to evaluate the \textit{Initial Generation} phase (in Section 5.1) during the Technical Evaluation study (in Section 6.3). It is based on the \textbf{English} script of the first physics lecture.}
\label{fig:dialog_ex1}
\end{figure*}

\begin{figure*}[!h]
\centering 
\includegraphics[width=\textwidth]{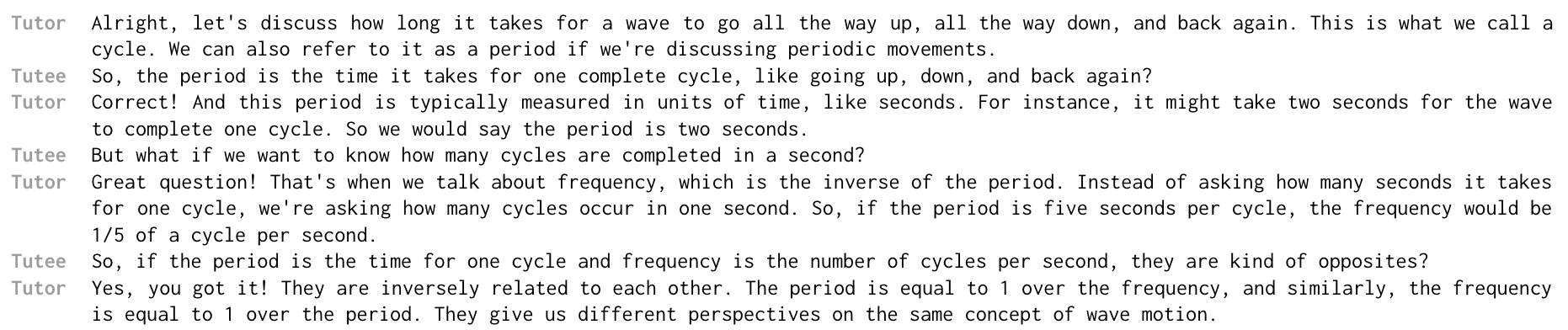} 
\caption{An example of technical evaluation dataset. This dialogue is one of the 16 examples generated by \textbf{Baseline} to evaluate the \textit{Initial Generation} phase (in Section 5.1) during the Technical Evaluation study (in Section 6.3). It is based on the \textbf{English} script of the first physics lecture.}
\label{fig:dialog_ex2}
\end{figure*}

Figure~\ref{fig:dialog_ex1} and Figure~\ref{fig:dialog_ex2} are two dialogue examples generated based on the English script of the physics lecture.

\subsection{Example 2}
\begin{figure*}[!h]
\centering 
\includegraphics[width=\textwidth]{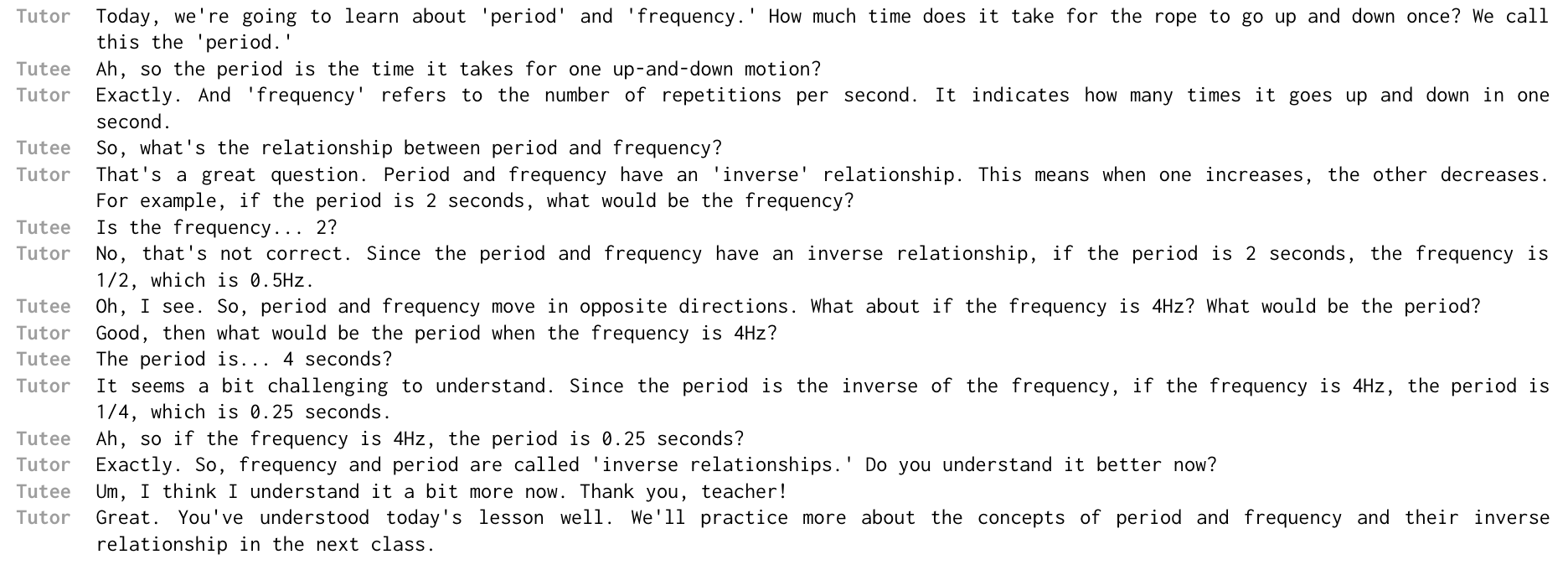} 
\caption{An example of technical evaluation dataset. This dialogue is one of the 16 examples generated by \textbf{\sysname{}} to evaluate the \textit{Initial Generation} phase (in Section 5.1) during the Technical Evaluation study (in Section 6.3). It is based on the Korean script of the first physics lecture. Thus, this example is a translation of a dialogue generated in \textbf{Korean}.}
\label{fig:dialog_ex3}
\end{figure*}

\begin{figure*}[!h]
\centering 
\includegraphics[width=\textwidth]{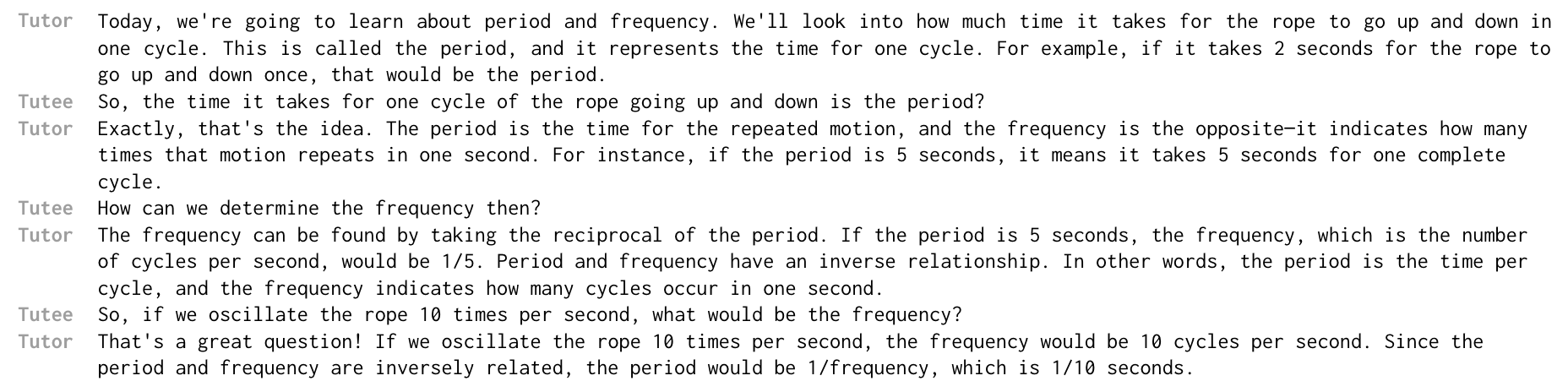} 
\caption{An example of technical evaluation dataset. This dialogue is one of the 16 examples generated by \textbf{Baseline} to evaluate the \textit{Initial Generation} phase (in Section 5.1) during the Technical Evaluation study (in Section 6.3). It is based on the Korean script of the first physics lecture. Thus, this example is a translation of a dialogue generated in \textbf{Korean}.}
\label{fig:dialog_ex4}
\end{figure*}

Figure~\ref{fig:dialog_ex3} and Figure~\ref{fig:dialog_ex4} are two dialogue examples generated from the Korean script of the physics lecture. We translated them into English as the output was Korean.

\end{document}